\documentclass{article}

\usepackage[a4paper,top=2cm,bottom=2cm,left=2cm,right=2cm]{geometry}
\usepackage{amsmath} 
\usepackage{amsfonts}
\usepackage{amssymb}
\usepackage{amsthm}
\usepackage{mathrsfs}
\usepackage{stackrel}
\usepackage[small,justification=justified]{caption}
\usepackage{authblk}
\usepackage{paralist}
\usepackage{graphics}
\usepackage{epsfig} 
\usepackage{graphicx}
\graphicspath{{./}{figures/}}
\usepackage{epstopdf}
\usepackage[colorlinks=true,linkcolor=blue,citecolor=red]{hyperref}
\usepackage{csquotes}

\newcommand{\bxi}{\boldsymbol{\xi}}
\newcommand{\x}{{\bf x}}
\newcommand{\bdelta}{{\boldsymbol{\delta}}}

\newtheorem{assum}{Assumption}

\providecommand{\keywords}[1]{\textbf{\textit{Keywords ---}} #1}

\newenvironment{sistem}
{\left\lbrace\begin{array}{@{}l@{}}}
{\end{array}\right.}

\begin{document}
\title{Kinetic and macroscopic equations for action potential in neural networks}
\author[1]{Martina Conte}
\author[2]{Maria Groppi}
\author[1]{Andrea Tosin\thanks{Corresponding author: \texttt{andrea.tosin@polito.it}}}
\affil[1]{\centerline{\small Department of Mathematical Sciences "G. L. Lagrange", Politecnico di Torino} \newline \centerline{\small Corso Duca degli Abruzzi 24 - 10129 Torino, Italy}}
\affil[2]{\centerline{\small Department of Mathematical, Physical, and Computer Science, University of Parma} \newline \centerline{\small Parco Area delle Scienze 53/A - 43124 Parma, Italy}}
\date{\today}                     
\setcounter{Maxaffil}{0}
\renewcommand\Affilfont{\itshape\small}
\maketitle

\begin{abstract}
Starting from the concept of binary interactions between pairs of particles, a kinetic framework for the description of the action potential dynamics on a neural network is proposed. It consists of two coupled levels: the description of a single brain region dynamics and the interactions among different regions. On one side, the pairwise interaction between neurons exchanging membrane potential is statistically described to account for the unmanageable number of neuron synapses within a single brain region. On the other, the network connections accounting for the brain region topology are represented and studied using concepts of the graph theory. Equilibrium and stability of the obtained macroscopic systems are analyzed as well as numerical simulations of the system dynamics are performed in different scenarios. In particular, the latter allows us to observe the influence of the discrete network topology on the membrane potential propagation and synchronization through the different regions, in terms of its spiking characteristics. 
\end{abstract}


\section{Introduction}\label{sec:intro}  
Collective behaviors and network dynamics have proven to play a key role in the modeling of complex systems in various fields. In particular, their importance is recognized in the mathematical description of information processing in the brain, both at macroscopic and cellular levels, and in related studies of synchronization phenomena. It is conjectured that synchronous brain activity is the most likely mechanism for many cognitive functions, such as attention and feature binding, as well as learning and memory formation. However, synchronization is not always useful, because brain disorders, such as schizophrenia, epilepsy, Alzheimer’s and Parkinson’s diseases, have been linked to high levels of synchronization in the neuronal activities \cite{jalili,bonacini2016single}.
 
Emerging collective behaviors like synchronization are strongly influenced by the microscopic interactions of the individual agents involved in the process. For this reason, in recent years, methodological approaches rooted in the statistical mechanics have gained a lot of momentum in the modeling and analysis of biophysical phenomena. Among them, we refer in particular to those which revisit methods of the classical kinetic theory to tackle systems that, at the microscopic scale, may be described as large ensembles of interacting particles. It is worth pointing out that, while in classical kinetic theory the word ``particle'' refers essentially to a gas molecule, in the modern kinetic theory for multi-agent systems it may indicate any kind of fundamental microscopic entity, whose interactions shape the dynamics of the entire system at hand. For instance, in this paper, a particle will be a brain cell, i.e., a neuron. The reason why the kinetic theory may be profitably exploited to investigate mathematically biophysical phenomena is that it provides a sound and flexible framework to bridge the often stochastic particle level, where elementary cell dynamics take place, and the aggregate level of tissue or organ dynamics. Without pretending to be exhaustive, we mention that, in the recent literature, examples of the use of kinetic theory methods for biophysical applications are found e.g. in the modeling of the onset and propagation of Alzheimer's disease~\cite{bertsch2017JPA,bertsch2021BM} as well as of tumor growth and therapeutic control strategies~\cite{preziosi2021JTB}.

In addition to the revisitation of kinetic theory methods, the contemporary literature about interacting multi-agent systems is increasingly focussed on \textit{networked} interactions. This means that, unlike gas molecules, the interactions among the considered particles are mediated by a background \textit{graph structure} discriminating which interactions may actually take place and at which rate. Examples of kinetic descriptions of networked interactions can be found, for instance, in opinion formation problems~\cite{burger2021VJM,loy2022PTRSA,toscani2018PRE} or in the modeling of the spread of infectious diseases~\cite{loy2021MBE}. In these problems, the background network may enter the big picture in essentially two ways. On one hand, particles themselves may be considered as the vertices of a usually large graph and interactions may take place only among particles directly connected by an edge of the graph. In this case, a \textit{statistical} description of the graph topology, viz. of the connections among the particles, is typically necessary, due to the large number of particles and connections which would make the detailed modeling of the graph structure unfeasible. On the other hand, particles may populate the vertices of a relatively small graph, which describes the structure of the connections among different particle communities. Within each node particles typically experience ``all-to-all'' interactions, namely interactions not mediated by a network structure, whereas interactions among particles located in different nodes are possible only if an edge exists between the corresponding nodes. Alternatively, interactions may be limited to particles belonging to the same node but particles may migrate across the nodes following the edges of the graph. In this case, a statistical description of the graph is not necessary due to the small number of nodes and vertices, which are then modeled one by one in detail.

In this paper, we adopt the methodological ideas just outlined to address action potential dynamics in neural networks. In particular, we aim to take into account both levels of the network description discussed above to build a model which, starting from simple neuron-to-neuron exchanges of electrical stimuli, is able to represent the activation/deactivation and synchronization of entire brain regions. For this, we combine a statistical description of the (large) network of neuron synapses within each brain region with a detailed description of the (small) network connecting the brain regions themselves. This allows us to recover, as a particular case, classical macroscopic neural network models featuring a coupling among brain regions described by a discrete Laplacian. In standard approaches to neural networks, these models are usually postulated directly at the macroscopic scale of the brain regions \cite{jalili,bonacini2016single}. Instead, with our approach, they are obtained consistently with elementary interactions among single neurons and can be possibly enriched with aggregate features resulting from the statistical description of the in-region network of neuron synapses. At a macroscopic level, we get at the end a complex network in which the nodes interact pairwise through a set of links, that encode the network topology. The evolution of the whole network is described by a system of coupled differential equations, and we investigate in this paper the complete synchronization of the resulting neural networks of dynamical systems. This is the simplest form of synchronization and consists in a perfect convergence of the trajectories of the single node systems in the course of time \cite{Boccaletti}, when some kind of coupling between them is introduced. The dynamical network is said to achieve globally (locally) asymptotic synchronization if the synchronous state, namely the solution in which the corresponding state variables are equal in each node, is globally (locally) asymptotically stable. The stability will be investigated by borrowing ideas from the Master Stability Function (MSF) approach \cite{pecora1998master}, taking advantage of the simple linear structure of the obtained network model. The MSF extends Lyapunov stability methods to complex networks of identical (or almost identical \cite{Sun}) oscillators and provides conditions for the local stability of the synchronous state, depending on the graph structure. 

The paper is organized as follows. In Section \ref{model}, we describe and derive the spatially homogenous model for the action potential dynamics in a single node and its generalization to the spatially inhomogeneous case, accounting for the partition of the brain into different macro-areas. Then, in Section \ref{microRule} we specify the microscopic rule of neuron-to-neuron interaction in the previously derived settings to describe the propagation of the action potential among neurons and we study the equilibria and their stability for both the spatially homogenous and inhomogeneous case. Section \ref{simulation} collects four sets of numerical tests aimed at showing the influence of the brain region topology on the propagation and synchronization of the action potential, its spiking characteristics, and its timing. Finally, in Section \ref{conclusion} we draw some conclusions and illustrate the relevance and the possible outcomes of this work.

\section{Model set up}\label{model}
 We propose a kinetic-based model for the description of the spread of an \enquote{information} on a neural network. We use the word \enquote{information} to  refer to any kind of mechanism whose dynamics are characterized by a neuron-to-neuron transmission. In this specific, we focus here on the propagation of the action potential among neurons. Considering the unmanageable number of nodes and links constituting the neuronal network that describes the connections among neurons, we propose a statistical description of the neuron connections. Precisely, to model the degree of connections of a single neuron we introduce the parameter $c$, which ranges between 0 and $+\infty$. Although this parameter $c$ should be described by a discrete variable, at this stage, we approximate it as a continuous one. This allows us to define the function $g(c)$, representing the distribution of the neuron connections, such that the quantity
\[
\mathbb{P}(c^*):=\int\limits_0^{c^*}g(c)dc 
\]

\noindent describes the probability that a neuron has a number of connections between 0 and $c^*$. \\
\indent In the following, we first assume that there is no spatial dependence on the function $g(c)$. This leads to a spatially homogenous model where each neuron has the same distribution of connections, independently of its spatial position. This choice is motivated by the fact the human brain can be divided into different macro-regions \cite{sporns2005human}, each of them controlling a particular neural activity and whose neurons have similar characteristics. Therefore, the corresponding spatially homogenous model would describe the neuron dynamics inside each macro-area. Since the connections among macro-areas have been well described \cite{bullmore2009complex,sporns2005human}, we then generalize the derived setting to the spatially inhomogeneous case, where $g=g(c,x)$. This second description allows us to take into account the variability of the distribution of neuron connections in the different brain regions. In this case, the system is properly described by a classical discrete network, where each node is a brain macro-area and each edge represents the connection between two macro-areas. Moreover, inside each node, the description of the connection would be given in statistical terms, following the spatially homogenous model, while the coupling terms between different macro-areas are deduced from the microscopic level. Indeed, we derive the macroscopic network model from a detailed description of the microscopic dynamics. This determines a novel justification and an original integration between a microscopic continuous and the macroscopic network description.

\subsection{Spatially homogenous case}\label{Homo_case}
\noindent Let $\mathcal{V}(t)$ be the dimensionless {\it membrane potential} of a single neuron \cite{izhikevich2007dynamical}, which we assume to evolve according to a neuron-to-neuron transmission mechanism, and $\mathcal{W}(t)$ the {\it recovery variable} associate to it. Precisely, the membrane potential is defined as the difference in electric potential between the interior and the exterior of a biological cell, depending on the concentration of different types of ions, among which sodium (Na$^+$) and potassium (K$^+$) play the major role. Its dynamics is driven by the opening of the Na$^+$ channels and the consequent emergence of a Na$^+$ electrochemical gradient.  A typical dynamics of the membrane potential is the so called {\it action potential}, which is defined as a sudden, fast, transitory, and propagating change of the resting membrane potential. Only neurons and muscle cells are capable of generating an action potential and this property is called the excitability. The recovery variable, instead, is introduced to acts as a negative feedback on the membrane potential, accounting for the activation of the K$^+$ ionic current and the inactivation of the Na$^+$ ionic current. Then, let $t\ge0$ be the time variable and $\mathcal{C}(t)$ the {\it neuron degree of connections}, the microscopic state of a single neuron is characterized by the triplet $(\mathcal{V},\mathcal{W},\mathcal{C})$ in the state space $[0, 1]\times[0,1]\times[0,+\infty)$. As such, the microscopic variables $\mathcal{V}$ and $\mathcal{W}$ have to be regarded as independent of $\mathcal{C}$, in the sense that their evolution is not a function of $\mathcal{C}$ and vice versa, and all variables are functions of time.
\begin{assum}
As it is reasonable to assume that no relevant changes of the cerebral tissue happen in the temporal window in which the action potential evolves, we assume that the number of connections $\mathcal{C}$ of a single neuron does not evolve in time. 
\end{assum}
\noindent In order to lighten the notation, we set $\mathcal{V}_t:=\mathcal{V}(t)$ and $\mathcal{W}_t:=\mathcal{W}(t)$. We rely on the concept of binary interactions between pairs of neurons to derive a Boltzmann-type mesoscopic model to describe the evolution of the microscopic state $(\mathcal{V}_t,\mathcal{W}_t,\mathcal{C})$  depending on repeated individual interactions with other neurons. Considering a generic pair of neurons with microscopic states $(\mathcal{V}_t,\mathcal{W}_t,\mathcal{C})$, $(\mathcal{V}^*_t,\mathcal{W}^*_t,\mathcal{C}^*) \in [0,1]\times[0,1]\times [0,+\infty)$, the effect of a binary interaction is modeled by 

$$L(\mathcal{V}_t,\mathcal{W}_t,\mathcal{V}^*_t,\mathcal{W}^*_t):[0, 1]^4\to[0, 1]^2$$ 
a prescribed function describing the interaction rule between the two pairs of neurons and accounting for the dynamics of the membrane potential and the recovery variable. To introduce a variation rule for $(\mathcal{V}_t,\mathcal{W}_t)$, we assume that in a short time interval $\Delta t>0$ there is a probability proportional to $\Delta t$ that the neuron undergoes the transmission mechanism. Precisely, we introduce the Bernoulli random variable $T_p\in\{0,1\}$ of parameter $p$, such that
\[
\mathbb{P}(\mathbb{T}_p=1)=p,\quad \mathbb{P}(\mathbb{T}_p=0)=1-p\,.
\]
This random variable describes whether an interaction leading to a change of microscopic state takes place ($\mathbb{T}_p=1$) or not ($\mathbb{T}_p=0$). To include the statistical description of the neuron connections, we assume that $p=\Delta t\, G(\mathcal{C},\mathcal{C}^*)$, where the function $G(\mathcal{C},\mathcal{C}^*)$ takes into account the probability that a neuron with microscopic states $(\mathcal{V},\mathcal{W},\mathcal{C})$ is connected by a synapsis with a neuron with microscopic state $(\mathcal{V}^*,\mathcal{W}^*,\mathcal{C}^*)$. The function $G(\mathcal{C},\mathcal{C}^*)$ influences the frequency of interactions in the collision operator. Several different expressions for the function $G$ can be proposed, each of them taking into account a different description of the probability of two neurons being connected. We derive the model in a general case without specifying the expression of $G$. Then, in the following sections, we would assume that $G(\mathcal{C},\mathcal{C}^*)=1$, indicating that all the neurons within a single brain region are connected. However, we remark that the general derivation here proposed allows for several different choices of $G$. 

\indent Considering $(\mathcal{V},\mathcal{W})$ the membrane potential related variables, the microscopic law for the pair $(\mathcal{V}_t,\mathcal{W}_t,\mathcal{C})$, $(\mathcal{V}^*_t,\mathcal{W}^*_t,\mathcal{C}^*)$ can be written as
\begin{equation}
\begin{sistem}
(\mathcal{V}_{t+\Delta t},\mathcal{W}_{t+\Delta t})=(1-\mathbb{T}_p)(\mathcal{V}_t,\mathcal{W}_t)+\mathbb{T}_p(\mathcal{V}'_t,\mathcal{W}'_t)\\[0.2cm]
(\mathcal{V}^*_{t+\Delta t},\mathcal{W}^*_{t+\Delta t})=(1-\mathbb{T}_p)(\mathcal{V}^{*}_t,\mathcal{W}^{*}_t)+\mathbb{T}_p(\mathcal{V}^{*'}_t,\mathcal{W}^{*'}_t)\,.\\
\end{sistem}
\end{equation}
We indicate with $(\mathcal{V}'_t,\mathcal{W}'_t):=(\mathcal{V}_t,\mathcal{W}_t)+L(\mathcal{V}_t,\mathcal{W}_t,\mathcal{V}^*_t,\mathcal{W}^*_t)$. To derive the Boltzmann-type equations for the evolution of $\mathcal{V}_t$ and $\mathcal{W}_t$, let us consider $\phi:[0,1]\times[0,1]\times[0,+\infty)\to \mathbb{R}$ a test function representing any observable quantity that can be computed out of the microscopic state $(\mathcal{V}_t,\mathcal{W}_t,\mathcal{C})$ of a neuron. It holds
\[
\phi(\mathcal{V}_{t+\Delta t},\mathcal{W}_{t+\Delta t},\mathcal{C})=\phi((\mathcal{V}_t,\mathcal{W}_t)+\mathbb{T}_pL(\mathcal{V}_t,\mathcal{W}_t,\mathcal{V}^*_t,\mathcal{W}^*_t),\mathcal{C})\,.
\]
For simplicity of notation in the following calculations, we set $\mathcal{A}:=(\mathcal{V},\mathcal{W})$. Considering the average of both sides and computing the mean with respect to $\mathbb{T}_p$ we get
\begin{equation}
\begin{split}
\langle\phi(\mathcal{A}_{t+\Delta t},\mathcal{C})\rangle&=\langle(1-p)\phi(\mathcal{A}_t,\mathcal{C})\rangle+\langle p\phi(\mathcal{A}_t',\mathcal{C})\rangle\\[0.2cm]
&=\langle\phi(\mathcal{A}_t,\mathcal{C})\rangle+\langle p\phi(L(\mathcal{A}_t, \mathcal{A}^*_t)+\mathcal{A}_t,\mathcal{C})\rangle-\langle p\phi(\mathcal{A}_t,\mathcal{C})\rangle\,.
\end{split}
\end{equation}
Recalling that $p=\Delta t \, G(\mathcal{C},\mathcal{C}^*)$, we can write
\begin{equation}
\langle\phi(\mathcal{A}_{t+\Delta t},\mathcal{C})\rangle=\langle\phi(\mathcal{A}_t,\mathcal{C})\rangle+\Delta t \, \langle G(\mathcal{C},\mathcal{C}^*)\phi(\mathcal{A}_t+L(\mathcal{A}_t, \mathcal{A}^*_t),\mathcal{C})\rangle-\Delta t \,\langle G(\mathcal{C},\mathcal{C}^*)\phi(\mathcal{A}_t,\mathcal{C})\rangle
\end{equation}
\begin{equation}
\Rightarrow \dfrac{\langle\phi(\mathcal{A}_{t+\Delta t},\mathcal{C})\rangle-\langle\phi(\mathcal{A}_t,\mathcal{C})\rangle}{\Delta t}=\langle G(\mathcal{C},\mathcal{C}^*)\left[\phi(\mathcal{A}_t+L(\mathcal{A}_t, \mathcal{A}^*_t),\mathcal{C})-\phi(\mathcal{A}_t,\mathcal{C})\right]\rangle\,.
\end{equation}
In the limit $\Delta t \to 0^+$, we get
\begin{equation}
\dfrac{d}{dt}\langle\phi(\mathcal{A}_{t},\mathcal{C})\rangle=\langle G(\mathcal{C},\mathcal{C}^*)\left[\phi(\mathcal{A}_t',\mathcal{C})- \phi(\mathcal{A}_t,\mathcal{C})\right]\rangle\,.
\end{equation}
Going back to the previous notation, we have
\begin{equation}
\dfrac{d}{dt}\langle\phi(\mathcal{V}_{t},\mathcal{W}_t,\mathcal{C})\rangle=\langle G(\mathcal{C},\mathcal{C}^*)\left[\phi(\mathcal{V}_t',\mathcal{W}_t',\mathcal{C})- \phi(\mathcal{V}_t,\mathcal{W}_t,\mathcal{C})\right]\rangle\,.
\label{limit_eq}
\end{equation}
Let us now define $f(v,w,c,t):[0,1]\times[0,1]\times[0,+\infty)\times\mathbb{R}_+\to\mathbb{R}$ the kinetic distribution function such that $f(v,w,c,t)dvdwdc$ represents the fraction of neurons that at time $t$ have membrane potential in $[v,v+dv]$, recovery variable in $[w,w+dw]$, and degree of connections in $[c,c+dc]$. We assume that the pair $(\mathcal{V}_t,\mathcal{W}_t,\mathcal{C})$, $(\mathcal{V}^*_t,\mathcal{W}^*_t,\mathcal{C}^*)$ consists in independent variables, such that their joint law is given by $f(v,w,c,t)f(v_*,w_*,c_*,t)$. We can, thus, rewrite \eqref{limit_eq} as 
\begin{equation}
\begin{split}
&\dfrac{d}{dt}\int\limits_0^1\int\limits_0^1\int\limits_0^{+\infty}\phi(v,w,c)f(v,w,c,t)dcdvdw\\[0.2cm]
&=\int\limits_0^1\int\limits_0^1\int\limits_0^1\int\limits_0^1\int\limits_0^{+\infty}\int\limits_0^{+\infty}G(c,c_*)\left[\phi(v',w',c)- \phi(v,w,c)\right]f(v,w,c,t)f(v_*,w_*,c_*,t)dcdc_*dvdv_*dwdw_*\,
\end{split}
\label{limit_eq2}
\end{equation}
where $(v',w'):=(v,w)+L(v,v_*,w,w_*)$ denotes the new state that a neuron with previous state $(v,w)$ acquires after an interaction with another neuron with state $(v_*,w_*)$. We assume that the number of connections is not affected by the interactions. Moreover, it has to hold
\[
\int\limits_0^1\int\limits_0^1f(v,w,c,t)dvdw=g(c)\quad \forall t>0\,.
\]
Considering equation \eqref{limit_eq2}, we can extract some information about the average macroscopic trends of our system. In particular, we can analyze the evolution of the average membrane potential $V(t)$ and the average recovery variable $W(t)$, defined as
\begin{equation}
V(t):=\int\limits_0^{+\infty}\int\limits_0^1\int\limits_0^1vf(v,w,c,t)dvdwdc\,,\quad W(t):=\int\limits_0^{+\infty}\int\limits_0^1\int\limits_0^1wf(v,w,c,t)dvdwdc\,
\end{equation}
respectively. Considering equation \eqref{limit_eq2} and choosing the test function as $\phi(v,w,c):=v \,\,\forall c\in[0,+\infty),\,\,\forall w\in[0,1]$, we get
\begin{equation}
\dfrac{dV(t)}{dt}=\int\limits_0^{+\infty}\int\limits_0^{+\infty}\int\limits_0^1\int\limits_0^1\int\limits_0^1\int\limits_0^1G(c,c_*)(v'-v)f(v,w,c,t)f(v_*,w_*,c_*,t)dvdv_*dwdw_*dcdc_*
\label{V_gen}
\end{equation}
while choosing $\phi(v,w,c):=w \,\,\forall c\in[0,+\infty),\,\,\forall v\in[0,1]$ we obtain
\begin{equation}
\dfrac{dW(t)}{dt}=\int\limits_0^{+\infty}\int\limits_0^{+\infty}\int\limits_0^1\int\limits_0^1\int\limits_0^1\int\limits_0^1G(c,c_*)(w'-w)f(v,w,c,t)f(v_*,w_*,c_*,t)dvdv_*dwdw_*dcdc_*\,.
\label{W_gen}
\end{equation}

\subsection{Spatially inhomogeneous case}\label{Inhom_case}
The above introduced spatially homogeneous model can be generalized to the spatially inhomogeneous case to account for the partition of the brain into different macro-areas, each of them  with a (possibly) different distribution of the neuronal connections \cite{bullmore2009complex}. Precisely, in this case, each neuron is identified by $(x,v,w,c)$, where $v$ indicates the membrane potential, $w$ the recovery variable, $c$ its degree of connections, and $x$ the macro-areas to which it belongs. In particular, $x$ is a discrete variable such that $x\in\{1,...,N\}=\mathbb{I}$, with $N$ the number of macro-areas. In this context, we define the distribution function $f:\mathbb{I}\times[0,1]\times[0,1]\times[0,+\infty)\times\mathbb{R}_+\to\mathbb{R}$ as a combination of the different distribution functions defined in each of the macro-areas, i.e., 
\begin{equation}\label{f_inhomo}
f=f(x,v,w,c,t):=\sum_{i=1}^Nf_i(v,w,c,t)\delta(x-i)\,.
\end{equation}
In this case, it holds
\[
\int\limits_0^1\int\limits_0^1f_i(v,w,c,t)dvdw=g_i(c)\quad \forall t>0\,\,\text{and} \,\,\forall i\in\mathbb{I}
\]
where $g_i(x)$ is the probability distribution of the neuronal connections in the $i$-th macro-area. The kinetic equation reads 
\begin{equation}
\begin{split}
&\dfrac{d}{dt}\int\limits_{\mathbb{I}}\int\limits_0^1\int\limits_0^1\int\limits_0^{+\infty}\phi(x,v,w,c)f(x,v,w,c,t)dcdwdvdx\\[0.3cm]
&=\iint\limits_{\mathbb{I}}\iint\limits_0^1\iint\limits_0^1\iint\limits_0^{+\infty}B(x,x_*)G(c,c_*)\left[\phi(x,v',w',c)- \phi(x,v,w,c)\right]\\[0.3cm]
&\hspace{3cm}f(x,v,w,c,t)f(x_*,v_*,w_*,c_*,t)dcdc_*dwdw_*dvdv_*dxdx_*\,.
\end{split}
\label{limit_eq_inhomog}
\end{equation}
Here, $B(x,x_*)G(c,c_*)$ represents the collision kernel. Even though, in general, the collision kernel can be given as a unique function of $(x,x_*,c,c_*)$, here we assume that it can be factorized in such a way that $B(x,x_*)$ describes the influence of the neuron location on the interactions, while $G(c,c_*)$ refers to their degree of connections. Substituting \eqref{f_inhomo} into \eqref{limit_eq_inhomog} we get 
\begin{equation}
\begin{split}
&\dfrac{d}{dt}\sum_{i=1}^N\int\limits_0^1\int\limits_0^1\int\limits_0^{+\infty}\phi_i(v,w,c)f_i(v,w,c,t)dcdwdv\\[0.3cm]
&=\sum_{i,j=1}^N\iint\limits_0^1\iint\limits_0^1\iint\limits_0^{+\infty}B_{ij}G(c,c_*)\left[\phi_i(v',w',c)- \phi_i(v,w,c)\right]f_i(v,w,c,t)f_j(v_*,w_*,c_*,t)dcdc_*dwdw_*dvdv_*
\end{split}
\end{equation}
where $\phi_i(v,w,c):=\phi(x=i,v,w,c)$ and $B_{ij}:=B(x=i,x_*=j)$. Fixing a certain index $i$ and assuming that $\phi(x,v,w,c)=\tilde{\phi}(x)\Phi(v,w,c)$, with $\tilde{\phi}(x)=1$ for $x=i$, while $\tilde{\phi}(x)=0$ for $x\ne i$, we get
\begin{equation}
\begin{split}
&\dfrac{d}{dt}\int\limits_0^1\int\limits_0^1\int\limits_0^{+\infty}\Phi(v,w,c)f_i(v,w,c,t)dcdwdv\\[0.3cm]
&=\sum_{j=1}^N\iint\limits_0^1\iint\limits_0^1\iint\limits_0^{+\infty}B_{ij}G(c,c_*)\left[\Phi(v',w',c)- \Phi(v,w,c)\right]f_i(v,w,c,t)f_j(v_*,w_*,c_*,t)dcdc_*dwdw_*dvdv_*\,.
\end{split}
\end{equation}
Defining the average membrane potential and recovery variables for the $i$-th macro-areas as
\begin{equation}
V_i(t):=\int\limits_0^{+\infty}\int\limits_0^1\int\limits_0^1vf_i(v,w,c,t)dvdwdc\,,\quad W_i(t):=\int\limits_0^{+\infty}\int\limits_0^1\int\limits_0^1wf_i(v,w,c,t)dvdwdc\,,
\end{equation}
respectively, we can derive the equations for the evolution of $V_i(t)$ and $W_i(t)$ in the spatially inhomogeneous case. Precisely, choosing $\Phi(v,w,c)=v$, we obtain
\begin{equation}
\dfrac{dV_i(t)}{dt}=\sum_{j=1}^N\int\limits_0^{+\infty}\int\limits_0^{+\infty}\int\limits_0^1\int\limits_0^1\int\limits_0^1\int\limits_0^1B_{ij}G(c,c_*)(v'-v)f_i(v,w,c,t)f_j(v_*,w_*,c_*,t)dvdv_*dwdw_*dcdc_*\,,
\label{V_gen_inhomo}
\end{equation}
while choosing $\Phi(v,w,c)=w$ we obtain
\begin{equation}
\dfrac{dW_i(t)}{dt}=\sum_{j=1}^N\int\limits_0^{+\infty}\int\limits_0^{+\infty}\int\limits_0^1\int\limits_0^1\int\limits_0^1\int\limits_0^1B_{ij}G(c,c_*)(w'-w)f_i(v,w,c,t)f_j(v_*,w_*,c_*,t)dvdv_*dwdw_*dcdc_*\,.
\label{W_gen_inhomo}
\end{equation}
In the following, we refer to this spatially inhomogeneous case as network system.

\section{Microscopic rule for action potential propagation}\label{microRule}
\noindent To derive the explicit equations for the average quantities $V(t)$ and $W(t)$, we need to specify the microscopic rule which would describe the propagation of the action potential between two neurons. Inspired by the nondimensional Morris–Lecar neuron model \cite{morris1981voltage} and the FitzHugh--Nagumo model \cite{fitzhugh1961impulses}, we define
\[
L(v,v_*,w,w_*):=(i_{ext}+\gamma(\bar{v}-v)+(v_*-v)-w,v-aw)
\]
with $i_{ext}$ representing all the external current sources stimulating the neuron, $\bar{v}$ the relaxation value for the membrane potential, while $\gamma>0$ and $a>0$ are dimensionless parameters. Here, the term $(v_*-v)$ is used to take into account at the microscopic level the coupling between two different neurons. Thus, the microscopic rule reads 
\begin{equation}\label{micro_rule}
\begin{sistem}
v'=v+i_{ext}+\gamma(\bar{v}-v)+(v_*-v) -w\\[0.2cm]
w'=w+v-aw
\end{sistem}
\end{equation} 
We recall that, depending on the choice of the function $G(c,c_*)$, it is possible to obtain different macroscopic spatially homogenous and inhomogeneous systems. As above stated, in these notes we choose $G(c,c_*)=1$. Thus, from equations \eqref{V_gen} and \eqref{W_gen}, for the spatially homogenous case we obtain the following system

\begin{equation}\label{macro_sis_hom}
\begin{sistem}
\dfrac{dV(t)}{dt}=i_{ext}+\gamma(\bar{v}-V(t))-W(t)\\[0.4cm]
\dfrac{dW(t)}{dt}=V(t)-aW(t)\,.
\end{sistem}
\end{equation}
Instead, from equations \eqref{V_gen_inhomo} and \eqref{W_gen_inhomo}, for the network system we obtain
\begin{equation}\label{macro_sis_inhom}
\begin{sistem}
\dfrac{dV_i(t)}{dt}=(i_{ext}+\gamma(\bar{v}-V_i(t))-W_i(t))\sum_{j=1}^NB_{ij}+\sum_{j=1}^NB_{ij}(V_j(t)-V_i(t))\qquad \forall i\in\mathbb{I}\\[0.4cm]
\dfrac{dW_i(t)}{dt}=(V_i(t)-aW_i(t))\sum_{j=1}^NB_{ij}\hspace{6.6cm} \forall i\in\mathbb{I}\,.
\end{sistem}
\end{equation}

\noindent We notice that system \eqref{macro_sis_inhom} represents a simpler form of the Morris--Lecar model defined on a graph and the derived term ${\sum_{j=1}^NB_{ij}(V_j(t)-V_i(t))}$ relates directly to the definition of the Laplacian on a graph. This term, typical of electrical synapses modeling \cite{bonacini2016single} and known as electric coupling, has not been added directly to the single-node system at the macroscopic level, but it has been derived from a more detailed definition of the microscopic dynamics. In particular, given the graph $\mathcal{G}=(\mathbb{I},\mathbb{E})$ containing a set $\mathbb{I}= \{1,2,...,N\}$ of nodes (brain macro-area) and a set $\mathbb{E}$ of arcs $(i,j)$ leading from initial node $i$ to terminal node $j$, we indicate with $B$ its adjacent matrix, taking into account the network topology into the system.  For $i=j$, we set $B_{ij}=1$, while for $i\ne j$ we set
\begin{equation}\label{laplacian}
B_{ij}=
\begin{sistem}
b_{ij}>0\quad\quad\,\,\,\text{if $(i,j)\in\mathbb{E}$} \\[0.2cm]
0\quad\,\quad\quad\quad\,\,\,\,\,\text{otherwise }\,.
\end{sistem}
\end{equation}
Recalling that an undirected graph is a graph where all the edges are bidirectional (namely, $B$ is symmetric), while with directed graph we refer to a graph in which the edges have a direction, we set $b_{ij}=1$ for undirected and directed graph (when $(i,j)\in\mathbb{E}$), while $b_{ij}$ could be lower or greater than $1$ if the graph is weighted, i.e., when each edge has an associated numerical value. 

\subsection{Equilibria and linear stability of the macroscopic systems}\label{staEq}
Here, we study equilibria and stability of the spatially homogenous \eqref{macro_sis_hom} and the network \eqref{macro_sis_inhom} systems. Starting from \eqref{macro_sis_hom}, we observe that the configuration
\begin{equation}\label{eq_homo}
\begin{sistem}
V^*=\dfrac{i_{ext}+\gamma\bar{v}}{\gamma+\dfrac{1}{a}}\\[1cm]
W^*=\dfrac{1}{a}V^*\,
\end{sistem}
\end{equation}
is the unique equilibrium of the single node system. We can prove that it is asymptotically stable by studying the eigenvalues of matrix $A$ associated with the single node dynamics, which is given by

\begin{equation*}
A=
\begin{pmatrix}
-\gamma & -1\\[0.2cm]
1 &-a
\end{pmatrix}\,.
\end{equation*}
We notice that $det(A)=\gamma a +1>0$, while $Tr(A)=-(\gamma+a)<0$, thus the real part of the corresponding eigenvalues is negative. This means that the equilibrium is asymptotically stable. Moreover, we observe that
\begin{equation}
\begin{split}
p(A-\lambda I)=0&\Longleftrightarrow (-\gamma-\lambda)(-a-\lambda)+1=0\\[0.3cm]
&\Longleftrightarrow \lambda^2+\lambda(\gamma +a)+1+a\gamma =0\\[0.1cm]
&\Longleftrightarrow \lambda_{1,2}=\dfrac{-(a+\gamma)\pm\sqrt{(a-\gamma)^2-4}}{2}\,.
\end{split}
\end{equation}
Thus,
 
$$\lambda_{1,2}\in\mathbb{R}\quad \Longleftrightarrow\quad  |a-\gamma|\ge 2\,.$$ If this condition is satisfied, the equilibrium is an asymptotically stable node, otherwise $\lambda_{1,2}\in\mathbb{C}$ and the equilibrium would be an asymptotically stable focus.\\
\indent Referring to the network system \eqref{macro_sis_inhom}, we can study its equilibria and stability using the framework of the Master Stability Function (MSF) approach, proposed in \cite{pecora1998master,Sun}. Precisely, starting from \eqref{macro_sis_inhom} we observe that
\[
\dfrac{dW_i(t)}{dt}=0\Longleftrightarrow (V_i(t)-aW_i(t))=0 \Longleftrightarrow W_i(t)=\dfrac{1}{a}V_i(t)\quad\,\forall i=1...N
\]
and

\begin{equation*}
\begin{split}
\dfrac{dV_i(t)}{dt}=0&\Longleftrightarrow (i_{ext}+\gamma(\bar{v}-V_i(t))-W_i(t))\sum_{j=1}^NB_{ij}+\sum_{j=1}^NB_{ij}(V_j(t)-V_i(t))=0\\[0.2cm]
&\Longleftrightarrow (i_{ext}+\gamma\bar{v}) \sum_{j=1}^NB_{ij}-\left(\gamma+\dfrac{1}{a}\right)V_i(t)\sum_{j=1}^NB_{ij}+\sum_{j=1}^NB_{ij}(V_j(t)-V_i(t))=0\\[0.2cm]
&\Longleftrightarrow V_i(t)=\dfrac{i_{ext}+\gamma\bar{v}}{\gamma+\dfrac{1}{a}}+\dfrac{\sum_{j=1}^NB_{ij}(V_j(t)-V_i(t))}{\left(\gamma+\dfrac{1}{a}\right)\sum_{j=1}^NB_{ij}}
\end{split}
\end{equation*}
Thus, the configuration
\begin{equation}\label{eq_inhomo}
\begin{sistem}
V^*_i=\dfrac{i_{ext}+\gamma\bar{v}}{\gamma+\dfrac{1}{a}}\quad\,\,\forall i=1...N\\[1cm]
W^*_i=\dfrac{1}{a}V^*_i\qquad\,\,\quad\,\forall i=1...N\,.
\end{sistem}
\end{equation}
is the synchronization manifold for our system. In particular we denote by $\x^*$ this synchronization manifold, whose $i$-th component $\x^*_i$ is the vector of the equilibrium state variables of the $i$-th region, i.e., $\x^*_i=(V^*_i,W^*_i)$, and it is given by

\begin{equation*}
\x_i^*=
\begin{pmatrix}
\dfrac{i_{ext}+\gamma\bar{v}}{\gamma+\dfrac{1}{a}}\\[1cm]
\dfrac{i_{ext}+\gamma\bar{v}}{a\gamma+1}
\end{pmatrix}\,\quad\,\forall i=1...N.
\end{equation*}
In order to use the MSF method for the stability analysis, we first rewrite system \eqref{macro_sis_inhom} using the classic notation of the graph theory as
\begin{equation}\label{gen_graphsys}
\dot{\x}_i(t)=F_i(\x_i)+M_i(\x_1,...,\x_N) \quad\,\forall i=1...N
\end{equation}
For each $i$-th node, $F_i(\x_i):=A_i\x_i+b_i$ is the vector field of single node dynamics, with

\begin{equation*}
A_{i}=B_i
\begin{pmatrix}
-\gamma & -1 \\[0.2cm]
1 & -a
\end{pmatrix}\,,
\quad b_i=B_i 
\begin{pmatrix}
i_{ext}+\gamma\bar{v}\\[0.2cm]
0
\end{pmatrix}
\end{equation*}
and $B_i=\sum_{j=1}^NB_{ij}$ degree of the $i$-th node. One difference with respect to the classical notation used in graph theory is that the vector field $F_i(\x_i)$ in this case depends on the $i$-th node. However, among the nodes this vector differs only for the multiplicative constant $B_i$, which does not influence the synchronization manifold or its stability. We are, thus, in the frame of a network of nearly identical dynamical systems \cite{Sun}. Instead, $M_i(\x_1,...,\x_N)$ is the vector field taking into account the coupling of $i$-th node with all the other nodes to which it is connected, i.e.,

\begin{equation*}
M_i(\x_1,...,\x_N)= 
\begin{pmatrix}
\sum_{j=1}^NB_{ij}(V_j-V_i)\\[0.2cm]
0
\end{pmatrix}.
\end{equation*}
We recall the definition of the Laplacian of the graph $L=(l_{ij})$ and the matrix for the diffusion profile $H$. Precisely, the Laplacian on the graph is defined as
\begin{equation}\label{laplacian}
L_{ij}=
\begin{sistem}
-B_{ij}\quad\,\,\text{for } \,i\ne j\\[0.2cm]
B_i\quad\,\quad\,\,\text{for }\, i=j
\end{sistem}
\end{equation} while, for the microscopic rule introduced in \eqref{micro_rule}, the matrix describing the diffusion profile reads

\begin{equation*}
H= 
\begin{pmatrix}
1 &0\\[0.2cm]
0 &0
\end{pmatrix}\,.
\end{equation*}
Thus, system \eqref{gen_graphsys} can be rewritten as
\begin{equation}
\dot{\x}_i(t)=F_i(\x_i)-\sum_{j=1}^NL_{ij}H\x_j \quad\,\,\forall i=1...N\,.
\end{equation}
To study the stability of the synchronization manifold $\x^*$, for each node we consider the perturbation ${\bdelta_i:=\x_i-\x^*_i}$ which satisfies the equation
\begin{equation}\label{delta_sys}
\dot{\bdelta}_i(t)=A_i\bdelta_i-\sum_{j=1}^NL_{ij}H\bdelta_j \quad\,\,\forall i=1...N\,.
\end{equation}
Considering the Laplacian matrix $L$, if it is diagonalizable in $\mathbb{R}$, then exists a matrix $T$ such that $TL=DT$, with $D$ diagonal matrix such that $D_{ii}=\epsilon_i$,$\forall i =1...N$, and $\epsilon_i\in\mathbb{R}$ eigenvalues of $L$. Thus, multiplying \eqref{delta_sys} by $T$ and defying $\bxi_i=\sum_{k=1}^NT_{ik}\bdelta_i^k$, we obtain the expression of system \eqref{delta_sys} in the canonical basis of the Laplacian eigenvalues
\begin{equation}\label{xi_sys}
\dot{\bxi}_i(t)=(A_i-\epsilon_iH)\bxi_i \quad\,\,\forall i=1...N\,,
\end{equation}
which represents the system of the normal modes. Here, the equilibrium $\bxi_i^*=\underline{0},\,\,\forall i=1..N,$ is asymptotically stable if and only if the eigenvalues of the associated matrix

\begin{equation*}
A_{i}-\epsilon_iH=B_i
\begin{pmatrix}
-\gamma-\dfrac{\epsilon_i}{B_i} & -1 \\[0.4cm]
1 & -a
\end{pmatrix}\,
\end{equation*}
have a negative real part. We observe that 
\begin{equation}
\begin{split}
&Tr(A_i-\epsilon_iH)=B_i(-a-\gamma-\dfrac{\epsilon_i}{B_i})<0\Longleftrightarrow \epsilon_i>-B_i(a+\gamma)\\[0.3cm]
&det(A_i-\epsilon_iH)=B_i^2\left(1+a\left(\gamma+\dfrac{\epsilon_i}{B_i}\right)\right)>0\Longleftrightarrow\epsilon_i>-B_i\left(\gamma+\dfrac{1}{a}\right)
\end{split}
\end{equation}
These two conditions can be written as 
\begin{equation}\label{cond_stab_MSF}
\epsilon_i>-B_i\min\left\{(a+\gamma), \left(\dfrac{1}{a}+\gamma\right)\right\}\,.
\end{equation}
Recalling that $\epsilon_i$ are the eigenvalues of the Laplacian matrix $L$, we can observe that this condition is always satisfied in two particular cases. 
\begin{itemize}
\item If we consider an undirected graph (weighted or not), then both $B$ and $L$ are symmetric. Therefore $L$ is diagonalizable with real and non-negative eigenvalues, i.e., $\epsilon_i\in\mathbb{R}$ and $\epsilon_i>0$ $\forall i=1..N$ and, thus, condition \eqref{cond_stab_MSF} is always satisfied and $\x^*$ is an asymptotically stable manifold for system \eqref{macro_sis_inhom}. It is worth noticing that the electrical coupling is bidirectional by construction, and this leads to biophysically meaningful undirected graphs for the network topology \cite{bonacini2016single}.
\item If $B$ is a triangular matrix, then also $L$ would have the same structure. Thus, it is diagonalizable with real and non-negative eigenvalues that allow to always satisfy condition \eqref{cond_stab_MSF}. This would be, for instance, the case of any directed graph for which $B_{ij}=0$ for any $i>j$. 
\end{itemize}
For a generic directed graph, it is not possible to obtain the same conclusion and the stability of the synchronization manifold would depend on the topology of the network described by $B$. As a final observation, we notice that, in the case of system \eqref{macro_sis_inhom}, the condition that distinguishes the case of $\x^*$ to be a node or a focus, meaning the eigenvalues of $(A_{i}-\epsilon_iH)$ being real, reads

 $$\left|a-\gamma-\dfrac{\epsilon_i}{B_i}\right|> 2\,.$$

\section{Numerical simulations}
\label{simulation}
\begin{figure}[!h]
     \centering
  \includegraphics[width=0.4\textwidth]{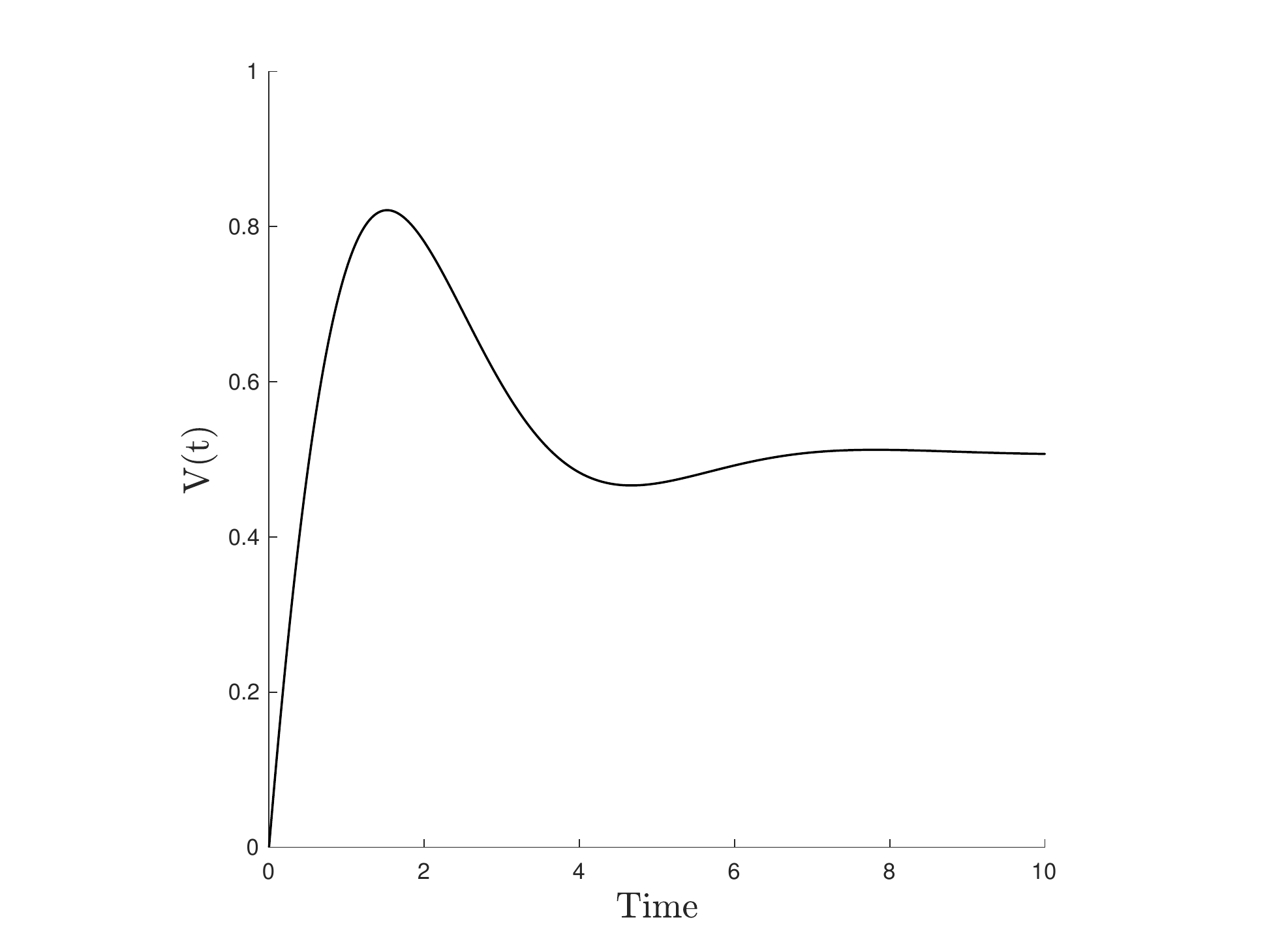}
\caption{Dynamics of the membrane potential variable $V(t)$ emerging from the system \eqref{macro_sis_hom}.}
    \label{SingleSrea}
\end{figure}

In this section, we present some numerical tests obtained by numerical integration of both the spatially homogenous model \eqref{V_gen}-\eqref{W_gen} with the microscopic rule \eqref{micro_rule}, summarized in  \eqref{macro_sis_hom}, and the corresponding network system \eqref{V_gen_inhomo}-\eqref{W_gen_inhomo} with the microscopic rule \eqref{micro_rule}, summarized in \eqref{macro_sis_inhom}. For the numerical simulation, we use a fourth order Runge--Kutta method. The simulation of the spatially homogenous model is used for a matter of comparison to analyze how the underlying network introduced in the spatially inhomogeneous model affects the single node dynamics, in terms of timing of the spike, amplitude, and width. Setting  $\bar{v}=1$, $\gamma=0.7$, $a=0.6$, and $i_{ext}=0.5$, in Figure \ref{SingleSrea} we show the dynamics of the action potential in a single macro-area for the initial condition $V(0)=0$ and $W(0)=0$.

\noindent This result shows how the developed model is able to replicate the emergence of a spike in the evolution of the action potential, representing the depolarizing and repolarizing phases of the considered area, as well as the undershoot phenomenon and the convergence towards the resting state. From the observation done in Section \ref{staEq} and considering the value of the parameters, we know that the equilibrium is an asymptotically stable focus. \\
\indent To show the capability of our approach to capture a large variety of scenarios, we shall present four main numerical tests performed with the network system \eqref{macro_sis_inhom}. \begin{itemize}
\item[{\bf Test 1:}] in Section \ref{noOrnoWe}, we consider undirected and non-weighted graphs. We first analyze the case of a complete undirected graph, in order to compare the effects of this topology with respect to the single node dynamics. Then, we analyze a configuration built on five nodes differently connected, studying the effect of the underlying network on the spiking characteristics of each region with respect to the others.
\item[{\bf Test 2:}] in Section \ref{OrnoWe}, we specify the previously discussed example for a directed and non-weighted graph, looking at the differences between the undirected case and the role of the network in this directed example. In particular, this test is motivated by the fact that there can be two type of communication between neurons, meaning electrical connections, which is bidirectional, and excitatory/inhibitory connections, which is unidirectional, and a directed network allows us to model this second scenario.
\item[{\bf Test 3:}] in Section \ref{noOrWe}, we specify the previous example to the case of an undirected and weighted graph. In particular, this test is motivated by the fact the physical distance between brain areas has an impact on the propagation of signals, such as the action potential, and a weighted network could be used to take this aspect into account. 
\item [{\bf Test 4:}] in Section \ref{ring} we consider the special case of a ring network in both the undirected and non-weighted case, directed and non-weighted case, and undirected and weighted case. 
\end{itemize}

\subsection{Test 1: undirected and non-weighted graph}\label{noOrnoWe}
\begin{figure}[!h]
     \centering
  \includegraphics[width=0.39\textwidth]{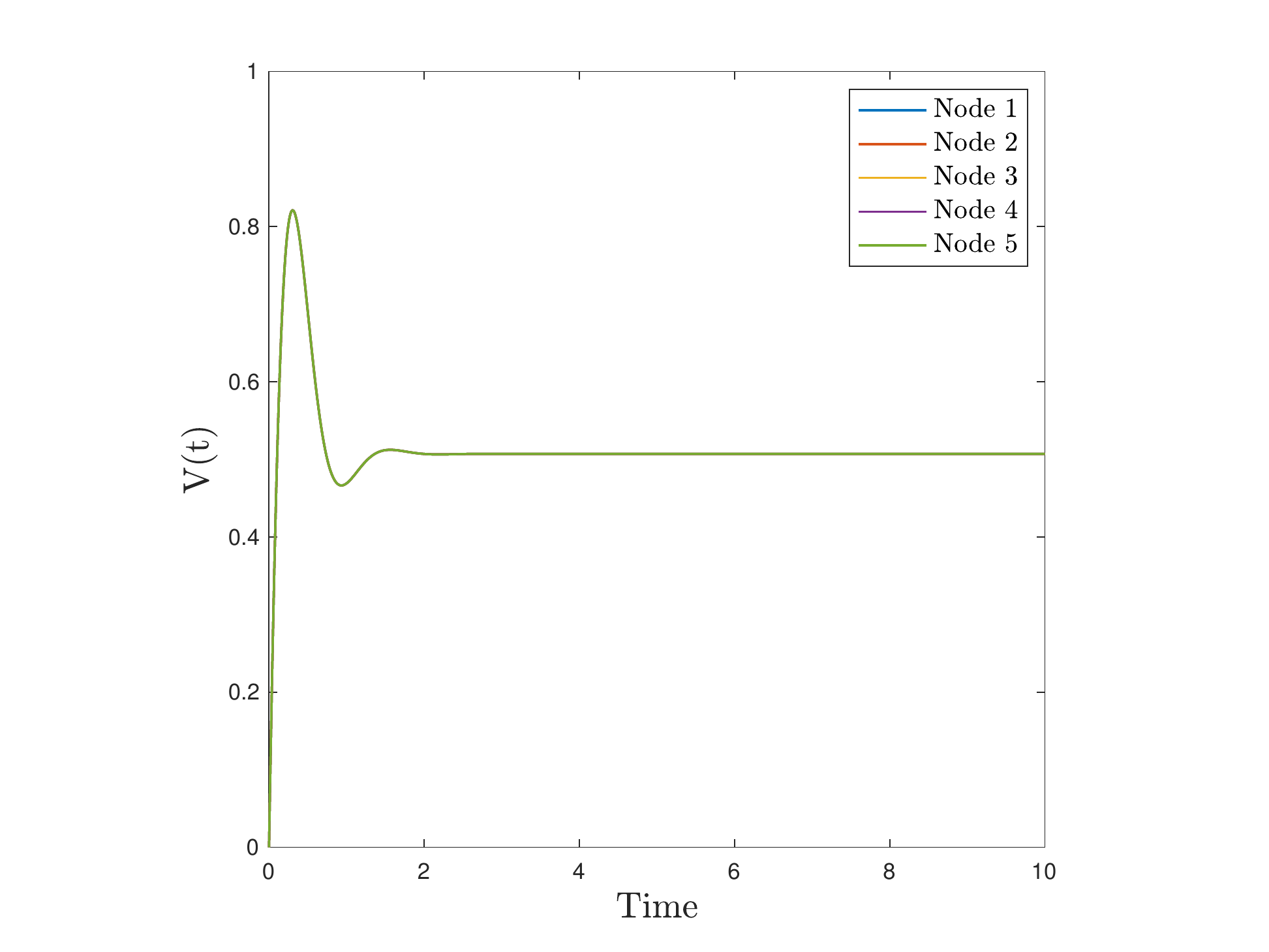}
    \includegraphics[width=0.39\textwidth]{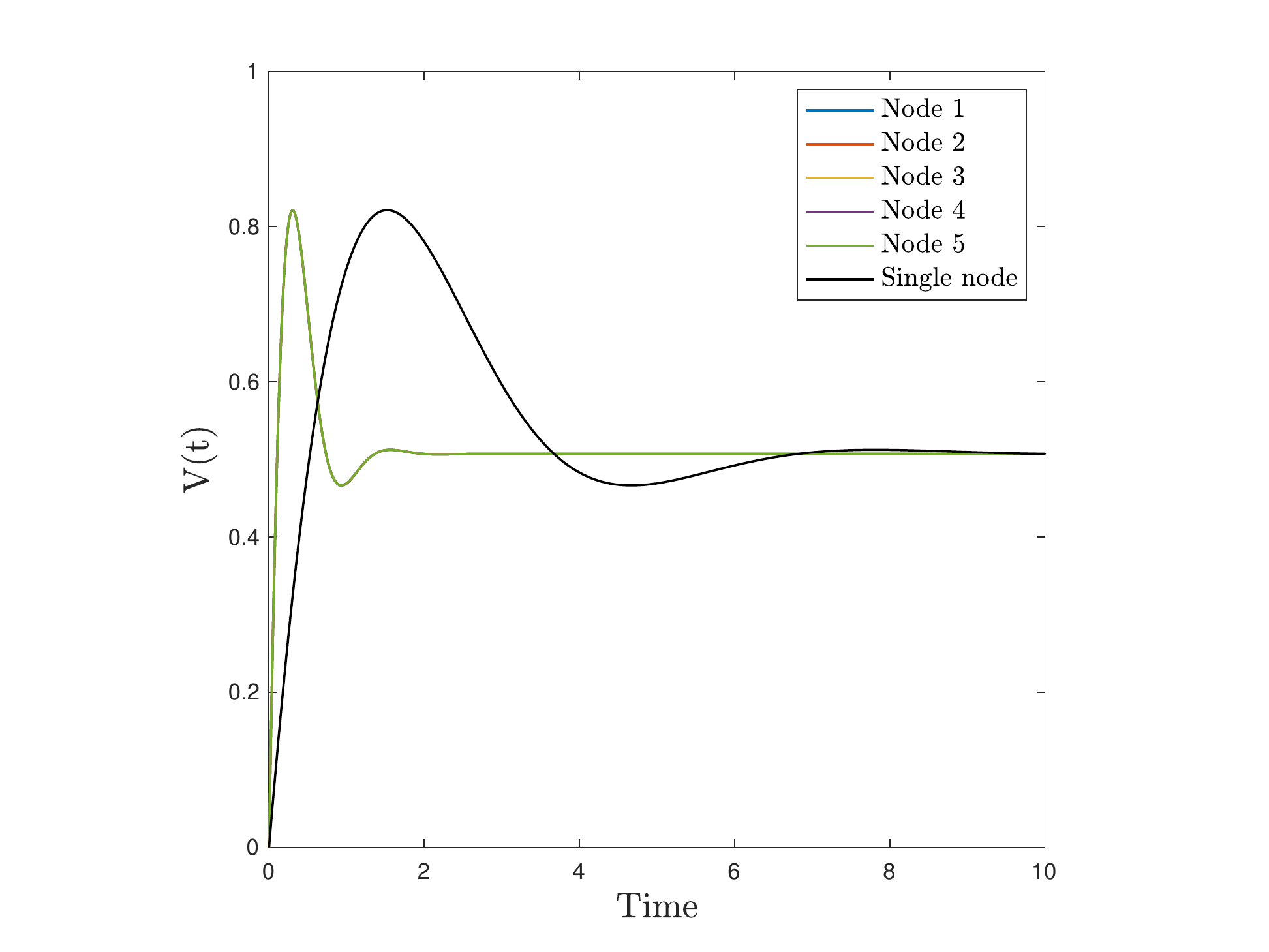}
\caption{{\bf Example 1.} Dynamics of the potentials $V_i(t)$, for $i=1..N$, emerging from system  \eqref{macro_sis_inhom} when a complete graph is considered (left plot). Comparison between the single node and the complete graph dynamics (right plot).}
    \label{Graph_ex1}
\end{figure}

We analyze here two examples of an undirected and non-weighted graph consisting of $N=5$ macro-areas differently connected. The different configurations of nodes and edges are described through the adjacency  matrix $B_{ij}$, for $i,j=1...N$ and, for all tests, we set the initial conditions $V_i(t)=W_i(t)=0$ $\forall i=1...N$. Moreover, we set $i_{ext}=0.5$, $\bar{v}=1$, $\gamma=0.7$, and $a=0.6$, as done for the simulation of Figure \ref{SingleSrea}. From the results of Section \ref{staEq}, we already know that the equilibrium configuration is asymptotically stable, independently of the specific graph topology.\\
\indent Firstly, as {\bf Example 1}, we consider a complete graph, meaning that we set $B_{ij}=1$ for each $i,j=1...N$. In this case, the evolution of the potentials $V_i(t)$ is described by the left plot of Figure \ref{Graph_ex1}, while on the right plot we compare the dynamics of a single node and of the 5 nodes network.

\begin{figure}[!h]
     \centering
    \includegraphics[width=0.7\textwidth,trim={0 5cm 0 4cm},clip]{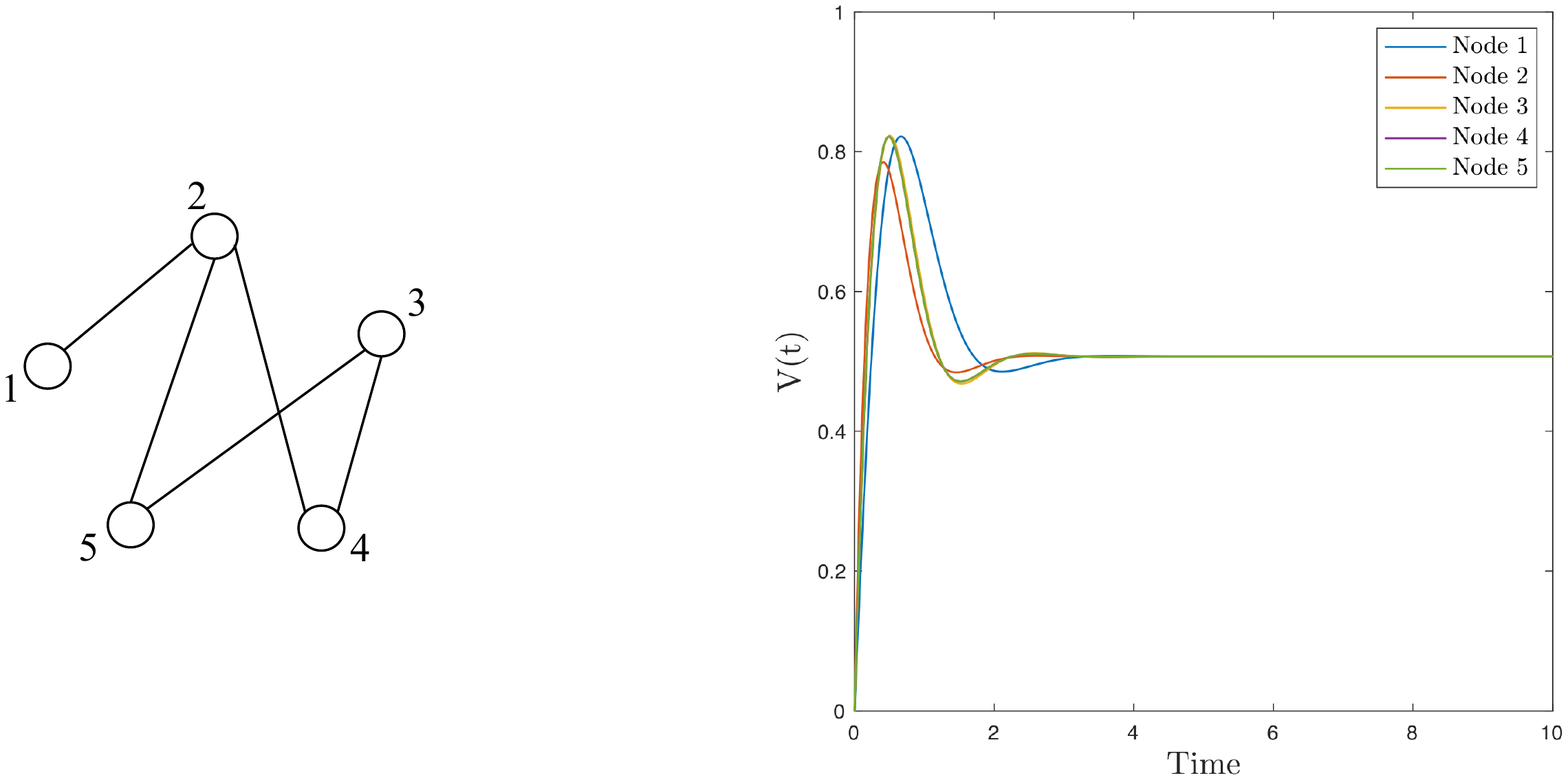}
\caption{{\bf Example 2.} Schematic representation of the graph described by the adjacent matrix \eqref{B_ex2} (left plot) and the dynamics of the potentials $V_i(t)$, for $i=1..N$, emerging from system \eqref{macro_sis_inhom} on this schematized network (right plot).}
    \label{Graph_ex2}
\end{figure}

\noindent From the left plot of Figure \ref{Graph_ex1}, we observe that considering a complete graph allows us to recover a perfect synchronization of the action potentials in all the nodes. The curves related to each of them are, in fact, perfectly overlapping and no differences can be grasped in either the spike time, width, or amplitude. Observing the comparison between the single node dynamics and the complete graph dynamics in the right plot, we notice that the connections between the different macro-areas determine a faster activation of the membrane potential and a shorter duration of the spike with respect to the case of an isolated node, i.e., a node not connected with any other. Therefore, this allows us to hypothesize that the undirected connections among nodes support action potential activation and its faster propagation in the network.\\
\indent As {\bf Example 2}, we consider the graph described by the following symmetric adjacency matrix 
\begin{equation}\label{B_ex2} 
B=
\begin{pmatrix}
1 &1& 0& 0& 0\\
1 &1 &0 &1 &1\\
0& 0& 1& 1& 1\\
0 &1 &1 &1 &0\\
0 &1 &1 &0 &1
\end{pmatrix}\,.
\end{equation}
The corresponding graph is schematized in the left plot of Figure \ref{Graph_ex2}, while the dynamics of the potentials $V_i(t)$ are shown in the right plot of Figure \ref{Graph_ex2}. 
\noindent With respect to the dynamics emerging in Figure \ref{Graph_ex1}, where a complete graph is considered, in Figure \ref{Graph_ex2} we observe that reducing the number of edges determines different delays in the action potential spike of each macro-area. Precisely, we can observe that the greater the number of connections of a node (meaning its degree), the earlier the spike occurs. In fact, node 2, which is the only one with three edges, shows the earliest spike; nodes 3, 4, and 5, which have the same degree, i.e., two edges, have almost identical dynamics with an intermediate spike emergence; node 1, instead, which has only one connection, is the last one to spike. Moreover, we can observe that, for the nodes with the same degree, the small differences in their action potential dynamics are determined by the characteristics of the node to which they are connected. In fact, nodes 4 and 5, which are connected with node 2 (that has the highest degree), have a slightly faster spike with respect to node 3.

\subsection{Test 2: directed and non-weighted graph}\label{OrnoWe}
\begin{figure}[!h]
     \centering
     \includegraphics[width=0.7\textwidth,trim={0 5cm 0 4cm},clip]{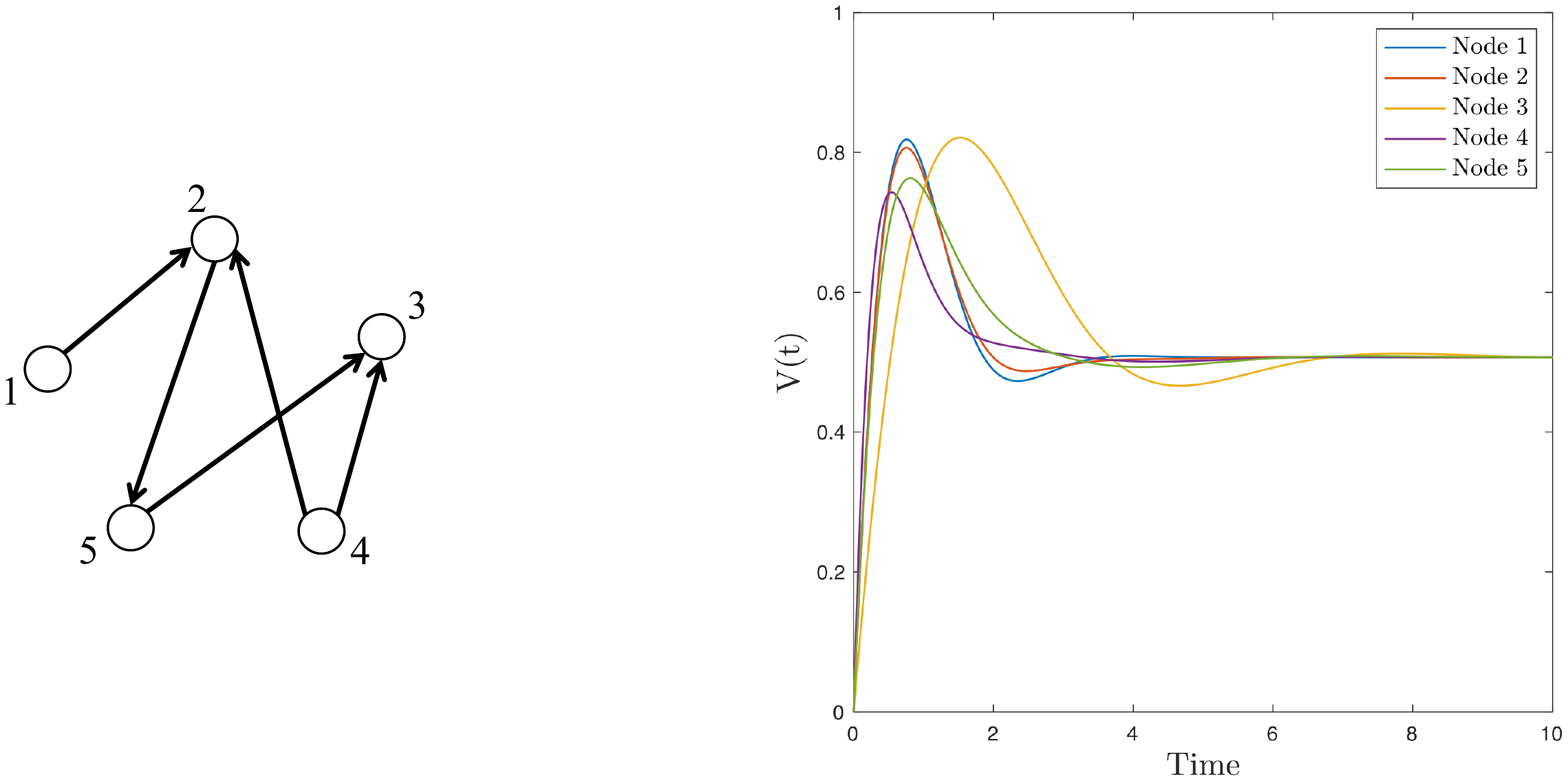}
\caption{{\bf Example 3.} Schematic representation of the graph described by the adjacent matrix \eqref{B_ex5} (left plot) and the dynamics of the potentials $V_i(t)$, for $i=1..N$, emerging from system \eqref{macro_sis_inhom} on this schematized network (right plot).}
    \label{Graph_ex5}
\end{figure}

Here, we analyze {\bf Example 3}, which describes a directed and non-weighted graph consisting of $N=5$ macro-areas differently connected.  Precisely, we consider the same configuration described in {\bf Example 2}, but choosing a specific orientation for each edge. The configuration of nodes and edges is described through the following (non symmetric) adjacency matrix
\begin{equation}\label{B_ex5} 
B=
\begin{pmatrix}
1 &1& 0&0& 0\\
0 &1 &0 &0 &1\\
0& 0& 1& 0& 0\\
0 &1 &1 &1 &0\\
0 &0 &1 &0 &1
\end{pmatrix}\,.
\end{equation}
The initial conditions are set to $V_i(t)=W_i(t)=0$ $\forall i=1...N$, and the constant parameters are the same of {\bf Test 1}. The graph is schematized in the left plot of Figure \ref{Graph_ex5}, while the dynamics of the  potentials $V_i(t)$ are shown in the right plot of the same Figure. 

\noindent Comparing the right plots of Figures \ref{Graph_ex2} and \ref{Graph_ex5}, we can immediately grasp the differences between the emerging dynamics. When a directed graph is considered, in fact, the timing at which each node spikes is related to the net outflow of a node, meaning to the differences between the number of outgoing edges and the number of incoming edges, rather than to its degree (as in the undirected case). If we indicate with ${O}_i$ the net outflow of the $i$-th node, it holds
\[
{O}_i=\sum_{j=1}^N (B_{ij}-B_{ji})\,.
\]
Precisely, the higher the value of this net outflow is, the earlier the spike occurs. For instance, we observe the dynamics of node 4, which has $O_4=2$, having two outgoing edges and no incoming edges, and of node 3, which has $O_3=-2$. Looking at the curves of the corresponding action potentials, node 4 shows the earliest spike, while node 3 is the latest. This example is clearly showing how much the network topology affects the system dynamics. In this case, we cannot use the results of Section \ref{staEq} to infer about the equilibrium stability. However, we can numerically calculate the eigenvalues of the associated Laplacian matrix, which results being $\epsilon_1=\epsilon_2=1$, $\epsilon_3=3$, $\epsilon_4=2$, and $\epsilon_5=3$. Thus, condition \eqref{cond_stab_MSF} is satisfied and the equilibrium manifold is asymptotically stable.

\subsection{Test 3: undirected and weighted graph}\label{noOrWe}
\begin{figure}[!h]
     \centering
     \includegraphics[width=0.7\textwidth,trim={0 5cm 0 4cm},clip]{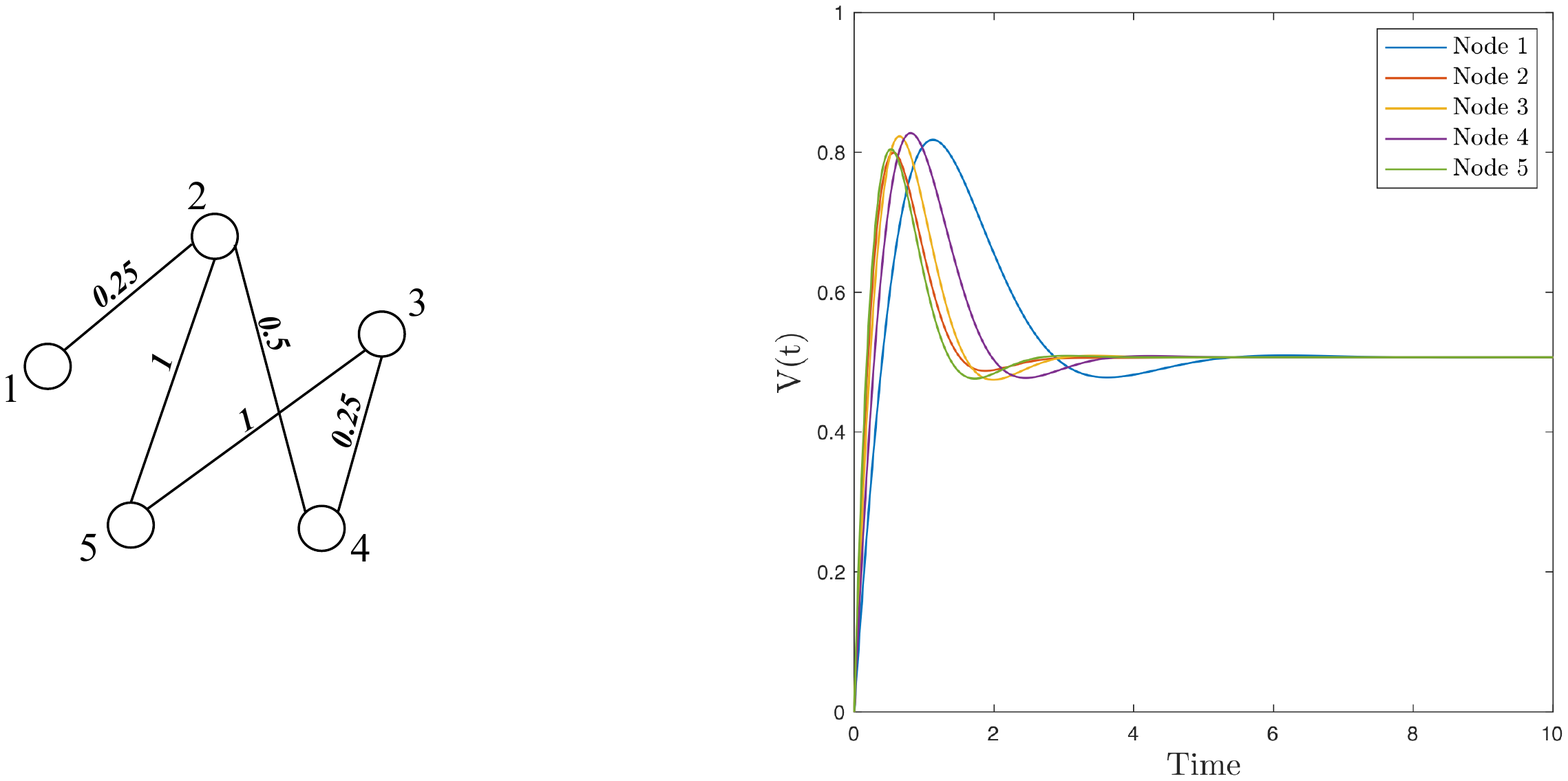}
\caption{{\bf Example 4.} Schematic representation of the graph described by the adjacent matrix \eqref{B_ex7} (left plot) and the dynamics of the potentials $V_i(t)$, for $i=1..N$ emerging from system \eqref{macro_sis_inhom} on this schematized network (right plot).}
    \label{Graph_ex7}
\end{figure}

For this test, we analyze the case of an undirected, but weighted graph consisting of $N=5$ macro-areas. Precisely, we consider the same configuration described in {\bf Example 2}, but choosing a specific weight for each edge. As for {\bf Test 1}, the analysis in Section \ref{staEq} allows us to directly conclude about the asymptotic stability of the equilibrium configuration, independently of the graph topology. The configuration of nodes and edges is described through the following symmetric adjacency matrix  
\begin{equation}\label{B_ex7} 
B=
\begin{pmatrix}
1 &0.25& 0& 0& 0\\
0.25 &1 &0 &0.5 &1\\
0& 0& 1& 0.25& 1\\
0 &0.5 &0.25 &1 &0\\
0 &1 &1 &0 &1
\end{pmatrix}\,.
\end{equation}
We refer to this as {\bf Example 4}, we set the initial conditions to $V_i(t)=W_i(t)=0$ $\forall i=1...N$ and we use the same constant parameters of {\bf Test 1} and {\bf 2}. In this case, the closer two macro-areas $i$ and $j$ are, the higher the value of $B_{ij}$ is. The corresponding graph is schematized in the left plot of Figure \ref{Graph_ex7}, while the dynamics of the potentials $V_i(t)$ are given in the right plot. 

\noindent If we assign to each node a value corresponding to the sum of the weights on its edges, i.e, for the $i$-th node we define the quantity
\[
T_i=\sum_{j=1}^NB_{ij}
\]
we can notice that the order of emergence of the spikes in Figure \ref{Graph_ex7} is directly related to $T_i$. Precisely, the greater $T_i$, the faster the $i$-th action potential spike.. For instance, if we look at node 1, the related value $T_1=0.25$ is the smallest with respect to the other nodes and, in fact, it is the last one to spike. Instead, node 5, which has $T_5=2$, shows the fastest dynamics and, in fact, it is the node with the biggest value of $T_i$. This phenomenon means that faster dynamics are promoted in those nodes that are more connected to the others and also closer to them, with respect to less connected and/or farther nodes.

\subsection{Test 4: ring graph}\label{ring}
\begin{figure}[!h]
     \centering
     \includegraphics[width=0.7\textwidth,trim={0 5cm 0 4cm},clip]{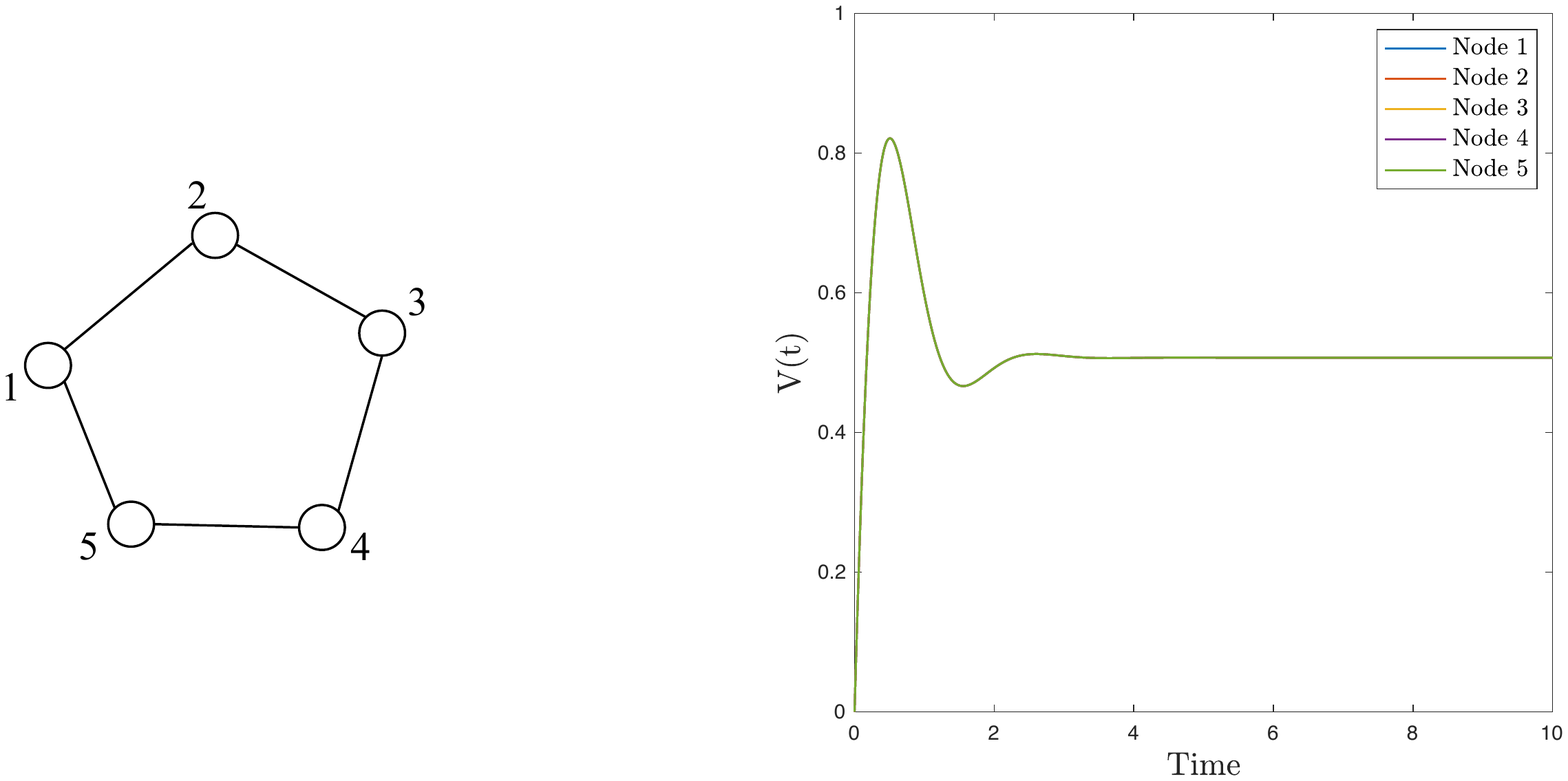}
\caption{{\bf Example 5.}  Schematic representation of the graph described  by the adjacent matrix \eqref{B_ex8} (left plot). Dynamics of the potentials $V_i(t)$, for $i=1..N$ emerging from system \eqref{macro_sis_inhom} (right plot).}
    \label{Graph_ex8}
\end{figure}

The last set of tests refers to the special case of a ring graph. We perform three sets of simulations, assuming the graph to be undirected ({\bf Example 5}), directed ({\bf Example 6}), or weighted ({\bf Example 7}), respectively. Starting from the undirected case, we consider the following symmetric adjacency matrix \begin{equation}\label{B_ex8} 
B=
\begin{pmatrix}
1 &1& 0&0& 1\\
1 &1 &1 &0 &0\\
0& 1& 1& 1& 0\\
0&0 &1 &1 &1\\
1 &0 &0 &1 &1
\end{pmatrix}
\end{equation}
A scheme of this graph is given in Figure \ref{Graph_ex8} (left plot), while the right plot shows the action potential dynamics.

\begin{figure}[!h]
     \centering
       \includegraphics[width=0.39\textwidth]{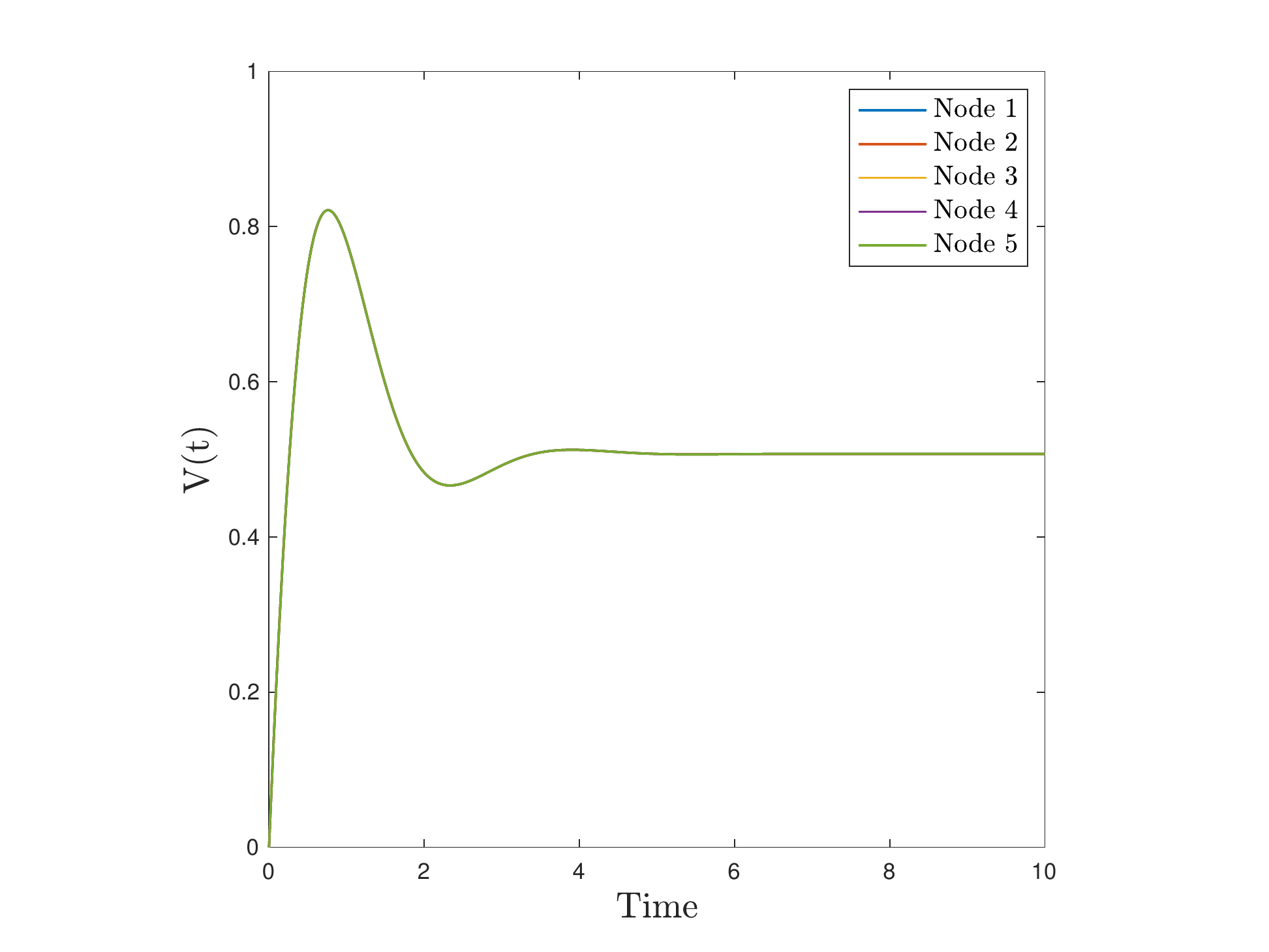}
  \includegraphics[width=0.39\textwidth]{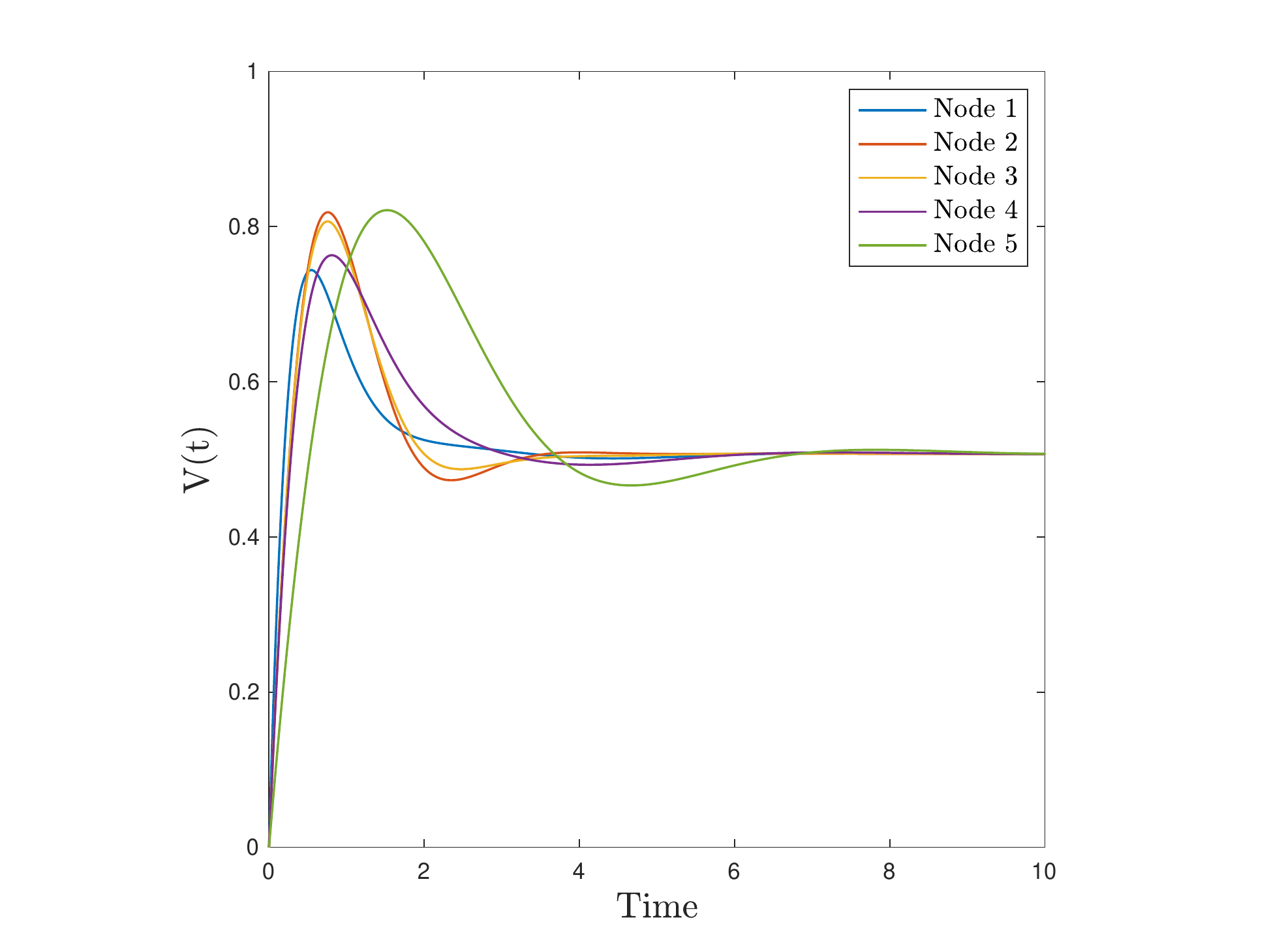}
\caption{{\bf Example 6.} Dynamics of the potentials $V_i(t)$, for $i=1..N$, emerging from system \eqref{macro_sis_inhom} when the connections are described by the adjacent matrix \eqref{B_ex9} with $B_{15}=0$ and $B_{51}=1$ (left plot) or with $B_{15}=1$ and $B_{51}=0$ (right plot).}
    \label{Graph_ex9}
\end{figure}

\noindent The particular topology of the ring net replicates the same behavior we observed in the case of a complete graph, i.e., full synchronization of the action potentials. In fact, in this case, all the nodes are identical and it is not possible to distinguish their dynamics. Moreover, because the graph is undirected, the asymptotic stability of its equilibrium configuration is guaranteed. \\
\indent Considering, instead, the case of a directed and non-weighted ring, we define two possible non-symmetric adjacency matrices, i.e.,
\begin{equation}\label{B_ex9} 
B=
\begin{pmatrix}
1 &1& 0&0& 0\\
0 &1 &1 &0 &0\\
0& 0& 1& 1& 0\\
0&0 &0 &1 &1\\
1 &0 &0 &0 &1
\end{pmatrix}
\end{equation}
and the analogous one in which we only change the direction of one edge, meaning $B_{15}=1$ and $B_{51}=0$. The dynamics of the  potentials $V_i(t)$ in these two cases are shown in Figure \ref{Graph_ex9}. Precisely, on the left plot we have the ring described by $B_{15}=0$ and $B_{51}=1$, while on the right plot $B_{15}=1$ and $B_{51}=0$.

\begin{figure}[!h]
     \centering
     \includegraphics[width=0.7\textwidth,trim={0 5cm 0 4cm},clip]{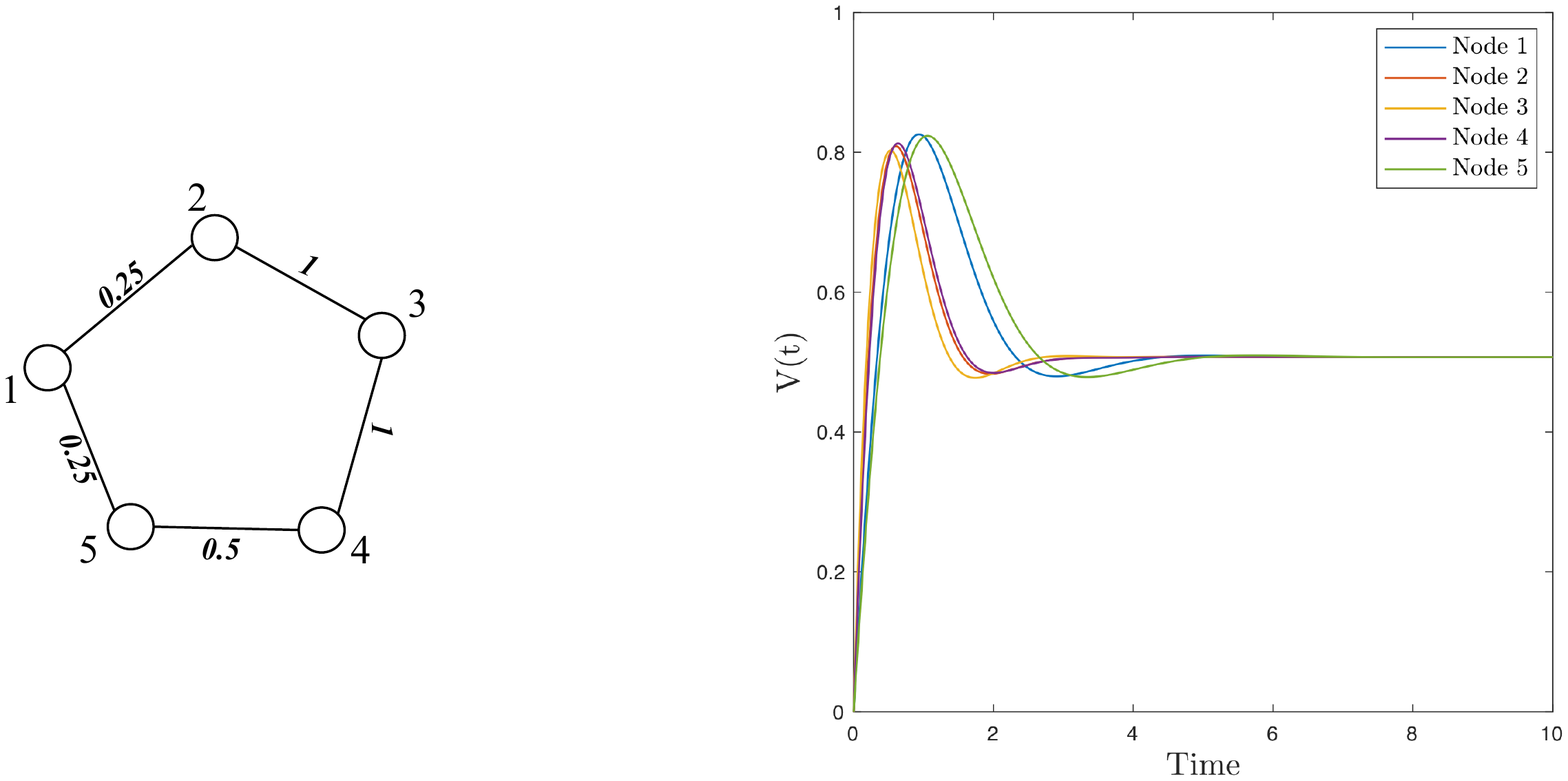}
\caption{{\bf Example 7.} Schematic representation of the graph described  by the adjacent matrix \eqref{B_ex10} (left plot) and the dynamics of the potentials $V_i(t)$, for $i=1..N$, emerging from system \eqref{macro_sis_inhom} on this schematized network (right plot).}
    \label{Graph_ex10}
\end{figure}

\noindent The left plot of Figure \ref{Graph_ex9} highlights the fact that, when each node has exactly one incoming and one outgoing edge, the resulting action potential dynamics does not show any difference with respect to the corresponding undirected graph. In fact, the nodes still remain indistinguishable.  Instead, changing the direction of at least one edge (as in the right plot of Figure \ref{Graph_ex9}), drastically modifies the evolution. Node 1, which in this case has $O_1=2$, has the fastest dynamics. Instead, node 5, which has $O_5=-2$, shows the slowest dynamics. The remaining nodes, which have $O_i=0$ for $i=2,3,4$, show the same timing for the spike, but different amplitudes. This effect could depend on the characteristics of the nodes to which they are connected: the more similar the connections of the adjacent nodes are, the closer  to their neighbors the node dynamics are. Concerning the stability of the equilibrium, when the adjacent matrix is described by the modified version \eqref{B_ex9} with $B_{15}=1$ and $B_{51}=0$, the Laplacian matrix is triangular and, thus, from the results in Section \ref{staEq}, the equilibrium manifold is asymptotically stable. Instead, when $B_{15}=0$ and $B_{51}=1$, we can numerically calculate the eigenvalues of the Laplacian matrix that result to be complex and such that the Laplacian matrix is diagonalizable in $\mathbb{C}$. From the numerical study of the matrix associated with the system of the normal mode, it is possible to prove the asymptotic stability of the equilibrium manifold.\\ 
\indent Finally, we consider the weighted and undirected ring described by the following symmetric adjacency matrix
\begin{equation}\label{B_ex10} 
B=
\begin{pmatrix}
1 &0.25& 0&0& 0.25\\
0.25 &1 &1 &0 &0\\
0& 1& 1& 1& 0\\
0&0 &1 &1 &0.5\\
0.25 &0 &0 &0.5 &1
\end{pmatrix}\,.
\end{equation}
The dynamics of the graph and a schematic representation of it are given in Figure \ref{Graph_ex10}.

\noindent As for the case of the directed graph in Figure \ref{Graph_ex9} (right plot), assigning weights to the edges perturbs the synchronized dynamics shown in Figure \ref{Graph_ex8} (right plot). Thus, even if the graph is undirected, instead of having a fully synchronized network, in the case of a weighted graph we observe delays among the spikes, which are related to the value $T_i$ assigned to each node. As for {\bf Example 4}, the stability of the equilibrium is guaranteed by the results of Section \ref{staEq}.

\section{Discussion}\label{conclusion}
In this note, we have studied the dynamics related to the spread of an information on a neural network, specializing the description to the case of the action potential propagation among neurons. On the basis of well-established studies on brain connectome \cite{sporns2005human}, we describe the human brain as a graph whose nodes correspond to the different brain regions and the edges to the connections among them, which can be obtained from tractography of diffusion tensor images \cite{sporns2005human}. The main focus of the proposed approach is the coupling between a local and statistical description of the interactions between neurons in a single brain region with a non-local and discrete description of the interactions between connected brain regions.\\
\indent  We analyze how the coupling affects the dynamics of the action potential at the macroscopic scale. Precisely, relying on the concept of binary interactions between pairs of neurons and starting from a microscopic description of them, we derive a Boltzmann-type mesoscopic equation in two specific scenarios. Firstly, considering a single brain region, we characterize each neuron by its number of connections and its membrane potential variables and we formally derived a spatially homogenous model from the microscopic description of neuron interactions. These are given in statistical terms using the function $g(c)$, which describes neuron connections. Then, we generalized the approach to the spatially inhomogeneous case, characterizing each neuron also by its discrete position in the brain connectome. This allows us to derive a macroscopic system for the average quantities describing the action potential evolution on the brain network. In particular, using this formalism we are able to obtain, in the equations at the macroscopic level, coupling terms of electric type among the different brain regions that correspond to the classical discrete Laplacian on a graph. However, here we do not add the coupling directly in the network system at the macroscopic level, but we derive it from a more precise description of the microscopic dynamics, thus providing a microscopic justification for its macroscopic expression. \\ 
\indent The specification of the microscopic interaction rule in a simple and linear case allows us to explicitly study equilibria and stability of the spatially homogenous and inhomogeneous models, which result to be linear, using techniques from the Master Stability Function approach. In particular, we obtain a general result about the asymptotic stability of the synchronization manifold of the network macroscopic system for a general undirected graph, independently of the specific network topology. Instead, we specify the condition describing the dependence of this stability on the network characteristics in the case of a directed graph. The numerical simulations proposed in the last sections reinforce our observations about the influence of the specific brain region connections on the overall dynamics, showing how they affect the propagation and synchronization of the action potential, its spiking characteristics, and its timing.\\
\indent 
As a first attempt, here we have chosen a simple expression for the function $G(c,c_*)$ describing the probability of two neurons being connected. However, several possible choices for this can be considered. Each of them would give rise to a different macroscopic model for the average membrane potential and recovery variable in a single brain region. Future plans will involve the use of different expressions for this function and the analysis of the corresponding system, which could result to be not closed and for which only estimations of the macroscopic behavior of $V(t)$ and $W(t)$ could be studied. At the same time, similar challenges in the rigorous derivation of the macroscopic equations could emerge when other choices for the microscopic rule are considered, some of which might involve a nonlinear coupling between the microscopic variables  \cite{izhikevich2004model}. Concerning the numerical results, we have shown here the most illustrative examples aimed at highlighting the main features of the influence of the network on action potential propagation. The choice of a simple network itself (five nodes differently connected) has been done with an illustrative purpose and a more realistic description of the neural network is planned to be included in future works, as well as the application of the proposed approach to the description of specific disease progression \cite{bertsch2017JPA,bertsch2017alzheimer}. Finally, we have mainly focused here on the synchronization equilibrium of the system and on the network conditions that can alter it. However, also the  asynchronous state might be of great interest. We have already obtained some preliminary results when an alternation of the input current on the different nodes is considered, and future plans include also a further analysis of asynchronization.

\section*{Acknowledgments}
The authors are members of INdAM-GNFM. The work has been performed in the frame of the Italian National Research Project "Integrated Mathematical Approaches to Socio--Epidemiological Dynamics" (Prin 2020JLWP23, CUP: E15F21005420006) and it was partially supported by the Ministry of University and Research through the MUR grant Dipartimento di Eccellenza 2018-2022, Project no. E11G18000350001. MC thanks the support of INdAM--GNFM Project "From kinetic to macroscopic models for tumor-immune system competition" (CUP$\_$E53C22001930001). The research of MG was granted by University of Parma through the action Bando di Ateneo 2022 per la ricerca co-funded by MUR-Italian Ministry of Universities and Research - D.M. 737/2021 - PNR - PNRR - NextGenerationEU” (project: "Collective and self-organised dynamics: kinetic and network approaches").

\bibliographystyle{plain}
\bibliography{CmGmTa-kinetic_action_potential}

\begin{thebibliography}{10}

\bibitem{bertsch2017alzheimer}
M.~Bertsch, B.~Franchi, N.~Marcello, M.C. Tesi, and A.~Tosin.
\newblock Alzheimer's disease: a mathematical model for onset and progression.
\newblock {\em Math. Med. Biol.}, 34(2):193--214, 2017.

\bibitem{bertsch2021BM}
M.~Bertsch, B.~Franchi, V.~Meschini, M.~C. Tesi, and A.~Tosin.
\newblock A sensitivity analysis of a mathematical model for the synergistic
  interplay of amyloid beta and tau on the dynamics of {A}lzheimer's disease.
\newblock {\em Brain Multiphysics}, 2:1--13, 2021.

\bibitem{bertsch2017JPA}
M.~Bertsch, B.~Franchi, M.~C. Tesi, and A.~Tosin.
\newblock Microscopic and macroscopic models for the onset and progression of
  {A}lzheimer's disease.
\newblock {\em J. Phys. A: Math. Theor.}, 50(41):1--22, 2017.

\bibitem{Boccaletti}
S.~Boccaletti, V.~Latora, Y.~Moreno, M.~Chavez, and D.U. Hwang.
\newblock Complex networks: structure and dynamics.
\newblock {\em Phys. Rep.}, 424(4):175--308, 2006.

\bibitem{bonacini2016single}
E.~Bonacini, R.~Burioni, M.~di~Volo, M.~Groppi, C.~Soresina, and A.~Vezzani.
\newblock How single node dynamics enhances synchronization in neural networks
  with electrical coupling.
\newblock {\em Chaos Solitons Fractals}, 85:32--43, 2016.

\bibitem{bullmore2009complex}
E.~Bullmore and O.~Sporns.
\newblock Complex brain networks: graph theoretical analysis of structural and
  functional systems.
\newblock {\em Nat. Rev. Neurosci.}, 10(3):186--198, 2009.

\bibitem{burger2021VJM}
M.~Burger.
\newblock Network structured kinetic models of social interactions.
\newblock {\em Vietnam J. Math.}, 49(3):937--956, 2021.

\bibitem{fitzhugh1961impulses}
R.~FitzHugh.
\newblock Impulses and physiological states in theoretical models of nerve
  membrane.
\newblock {\em Biophys. J.}, 1(6):445--466, 1961.

\bibitem{izhikevich2004model}
E.~M. Izhikevich.
\newblock Which model to use for cortical spiking neurons?
\newblock {\em IEEE trans. neural netw.}, 15(5):1063--1070, 2004.

\bibitem{izhikevich2007dynamical}
Eugene~M Izhikevich.
\newblock {\em Dynamical systems in neuroscience}.
\newblock MIT press, 2007.

\bibitem{jalili}
M.~Jalili.
\newblock Synchronizing {H}indmarsh-{R}ose neurons over {N}ewman–{W}atts
  networks.
\newblock {\em Chaos}, 19(3):033103, 2009.

\bibitem{loy2022PTRSA}
N.~Loy, M.~Raviola, and A.~Tosin.
\newblock Opinion polarization in social networks.
\newblock {\em Philos. Trans. Roy. Soc. A}, 380(2224):1--15, 2022.

\bibitem{loy2021MBE}
N.~Loy and A.~Tosin.
\newblock A viral load-based model for epidemic spread on spatial networks.
\newblock {\em Math. Biosci. Eng.}, 18(5):5635--5663, 2021.

\bibitem{morris1981voltage}
C.~Morris and H.~Lecar.
\newblock Voltage oscillations in the barnacle giant muscle fiber.
\newblock {\em Biophys. J.}, 35(1):193--213, 1981.

\bibitem{pecora1998master}
L.~M. Pecora and T.~L. Carroll.
\newblock Master stability functions for synchronized coupled systems.
\newblock {\em Phys. Rev. Lett.}, 80(10):2109, 1998.

\bibitem{preziosi2021JTB}
L.~Preziosi, G.~Toscani, and M.~Zanella.
\newblock Control of tumor growth distributions through kinetic methods.
\newblock {\em J. Theor. Biol.}, 514:110579, 2021.

\bibitem{sporns2005human}
O.~Sporns, G.~Tononi, and R.~K{\"o}tter.
\newblock The human connectome: a structural description of the human brain.
\newblock {\em PLoS Comput. Biol.}, 1(4):e42, 2005.

\bibitem{Sun}
J.~Sun, E.~M. Bollt, and T.~Nishikawa.
\newblock Master stability functions for coupled nearly identical dynamical
  systems.
\newblock {\em EPL}, 85(6):60011, 2009.

\bibitem{toscani2018PRE}
G.~Toscani, A.~Tosin, and M.~Zanella.
\newblock Opinion modeling on social media and marketing aspects.
\newblock {\em Phys. Rev. E}, 98(2):1--15, 2018.

\end{thebibliography}

\end{document}